\documentclass[final, twocolumn]{elsarticle}

\usepackage{graphicx}
\usepackage{tabularx}
\usepackage{multirow}
\usepackage{listings}
\usepackage{caption}
\usepackage[caption=false]{subfig}

\begin{document}
\begin{frontmatter}

\title{Smartphone-based crowdsourcing \\for estimating the bottleneck capacity in wireless networks}

\author[label1]{Enrico Gregori\fnref{alorder}}
\author[label1]{Alessandro Improta\fnref{alorder}}
\author[label2]{Luciano Lenzini\fnref{alorder}}
\author[label2]{Valerio Luconi\fnref{alorder}}
\author[label2]{Nilo Redini\fnref{alorder}}
\author[label2]{Alessio Vecchio\fnref{alorder}\corref{cor1}}
\ead{alessio.vecchio@unipi.it}

\address[label1]{IIT-CNR, Pisa, Italy}
\address[label2]{Dip. di Ingegneria dell'Informazione, Universit\`a di Pisa, Pisa, Italy}

\fntext[alorder]{Authors are listed in alphabetical order.}
\cortext[cor1]{Corresponding author at: Dip. di Ingegneria dell'Informazione,
Universit\`a di Pisa, Largo L. Lazzarino 1, 56122 Pisa, Italy.}

\begin{abstract}

Crowdsourcing enables the fine-grained characterization and performance
evaluation of today's large-scale networks using the power of the masses and
distributed intelligence. This paper presents SmartProbe, a system that
assesses the bottleneck capacity of Internet paths using smartphones, from a
mobile crowdsourcing perspective. With SmartProbe measurement activities are
more bandwidth efficient compared to similar systems, and a larger number of
users can be supported. An application based on SmartProbe is also presented:
georeferenced measurements are mapped and used to compare the performance of
mobile broadband operators in wide areas. Results from one year of operation
are included.

\end{abstract}


\end{frontmatter}

\section{Introduction}

Crowdsourcing exploits the help of the masses to solve large-scale problems. A
task that is too demanding for the internal resources of a single organization
can be divided into small and loosely-coupled activities which are assigned to
and carried out by a population of individuals~\cite{howe2008:crowd}. Unlike
outsourcing, with crowdsourcing the identity of the cooperating users is
generally not relevant, and the workforce dynamically changes according to the
necessities of the delegating organization and the will of the participants.
Crowdsourcing is used in a variety of contexts, from the production of creative
content to the massive analysis of data. Notable examples include
Threadless~\cite{threadless}, an online clothes shop where the design of items
is collaborative and user-driven, and SETI@home, a distributed effort aimed at
searching for extraterrestrial intelligence using spare CPU cycles of users'
machines~\cite{seti}.  There are also several platforms that support
crowdsourcing-based interaction, such as Amazon's Mechanical Turk~\cite{mturk},
Microworkers~\cite{microworkers}, and Crowdflower~\cite{crowdflower}. These platforms
provide other companies with methods for submitting tasks, enrolling users, and
managing payments. Although in some cases the activities delegated to users are
trivial, in other situations the tasks are complex and require intelligence
and/or creativeness. In all cases, the whole result is greater than the sum of
its parts, and decentralized and distributed intelligence, aggregated through
crowdsourcing, can provide an answer to unsolved scientific and
engineering problems.

Crowdsourcing systems have traditionally been based on the web, as it provides
collaboration tools that are both efficient and easy to use~\cite{doan11:crowdsourcing}.
More recently, the web-centric interaction model has been expanded to support
smartphone-based cooperation~\cite{chatz12:crowdsourcing}. In fact, smartphones are
an appealing platform for crowdsourcing applications: they are always on and
carried by their owners, they are mobile and richly connected, and they are
equipped with an increasing number of sensors (camera, microphone, etc). When using a
smartphone, the working user is no longer constrained to a fixed position and
he/she can carry out the requested task at different locations, possibly using
additional input mechanisms. The term \emph{crowdsensing} is used when sensing
is the prevalent activity delegated to participants.  Crowdsensing applications
can be classified according to the type of phenomenon being measured. Examples
include environmental applications (for observing pollution and water
levels), infrastructure monitoring applications (for collecting information
about traffic congestions and road conditions), and social applications (for
monitoring activity levels of
individuals)~\cite{ganti11:crowdsensing,Vivacqua2012189}.  In several
scenarios, the crowdsensing activities are carried out without a well-defined
employer-employee relationship. Users may be interested in participating in
sensing activities for a number of reasons, including altruism or the
scientific relevance of the end goals. In other cases they perceive the results
of the sensing activity as also being useful for themselves (even though the
small task each user completes may be scarcely significant, the level of
interest in the aggregated results can be much higher).

An increasing number of crowdsourcing/crowdsensing systems are related to
networking: the crowd-based approach provides a solution to the need for
collecting detailed information on today's large-scale networked systems.  For
instance crowdsourcing is currently used for the following purposes: $i)$
detection of traffic differentiation silently applied by Internet service
providers to their customers~\cite{glasnost}, $ii)$ characterization of the
Internet and detection of network problems~\cite{ssh05,sanchez13:dasu}, $iii)$
analysis and measurement of wireless networks~\cite{zee}.

This paper presents SmartProbe, a network measuring tool designed to operate in
a mobile crowdsensing scenario. Using smartphones as measuring elements,
SmartProbe estimates the bottleneck capacity of Internet paths. Since it is
executed on user devices, SmartProbe was designed and customized to generate
less traffic, and thus to use less energy, compared to similar systems
(experimental results show a significant reduction in sent/received data).  In
addition, since the system has to be used by a possibly large number of users,
the software infrastructure on the server-side was designed to cope with
multiple measurement requests.  Measurements can be georeferenced thanks
to the self-positioning ability of smartphones. A demo application is also
presented: different mobile broadband operators are compared using crowdsourced
measurements. Other possible uses include mapping the performance of Wi-Fi
access points in an urban area or analyzing the performance of a cell phone
operator in relation to variations in user positions.

\section{Background and motivation}

\begin{figure}[t]
   \centering
   \includegraphics[width=1.0\columnwidth]{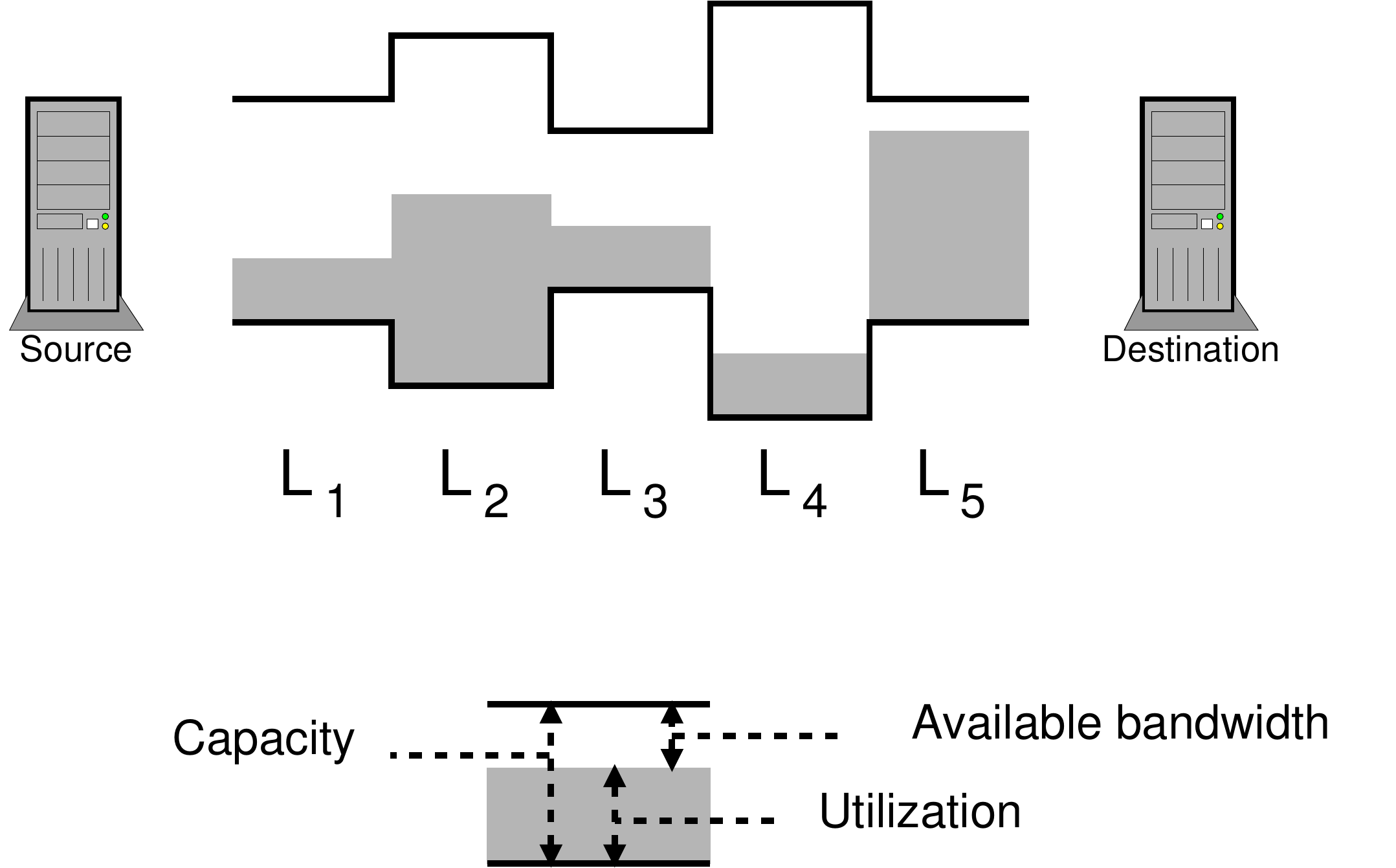}
   \caption{An Internet path with links characterized by different capacities and utilization levels.}
   \label{fig:path}
\end{figure}

Let $L_i$ be the $i$-th link of an Internet path, with $i \in {1..n}$.  The
capacity $C_i$ of the $i$-th link is the maximum data rate that link $L_i$ can
achieve.  The bottleneck capacity of an Internet path is defined as the
capacity of the narrowest link of the path considered~\cite{Claffy}, and thus
it is equal to $\min \{C_1, ..., C_n\}$ (in other words it is the capacity of
the link that imposes a bottleneck on the path in terms of data rate). The
capacity of a link (or a path) is a static property that does not change with
time, and should not be confused with the available bandwidth. The latter
represents the residual bandwidth not currently in use by other traffic (a
property whose value depends on current
conditions)~\cite{Jain:2004:TFP:1028788.1028825,Jain:2003:EAB:941374.941378}.
If $A_i(t)$ is the available bandwidth of $L_i$ at time $t$, then $C_i \ge
A_i(t)$ holds. For example, the bottleneck capacity of the path shown in
Figure~\ref{fig:path} is $C_3$, the capacity of $L_3$, whereas the link with
the smallest available bandwidth is $L_5$ (considering the current utilization
level as depicted by the gray areas).

The ability to measure the bottleneck capacity is useful not only for network
protocols (e.g. for congestion control) but also at the application and user
levels. In fact, this information can be used to tune the operational parameters
of streaming applications and peer-to-peer systems, or to evaluate the actual
performance of a residential ISP connection.  

Methods and techniques for measuring this network property have received
significant attention from both researchers and practitioners
~\cite{Dovrolis:2004:PTC:1046014.1046015,CapProbe,Chen}. In addition, while
most of the initial activities have been carried out for wired networks, more
recently the study of techniques specifically designed for wireless
environments has gained momentum~\cite{Guang,Li}.  In this paper we move
forwards in two different directions. On the one hand, we continue this trend
by giving even more relevance to wireless scenarios. The aim is to make the
evaluation of the bottleneck capacity more suitable for execution on mobile
devices (smartphones and tablets, which have surpassed common PCs in
terms of sales, now represent the preferred Internet-enabled devices for the
majority of users).  On the other hand, we believe that an evolution of
bottleneck capacity estimation tools in a crowdsourcing perspective may pave
the way for interesting and unexplored usages, as it may provide fine-grained
information on today's large scale networked  systems.

The evaluation of the bottleneck capacity in a smartphone-based crowdsourcing
scenario must take into account $i)$ bandwidth and energy efficiency, $ii)$
tuning for wireless connections, and $iii)$ support for possibly large numbers
of users.

\noindent {\bf Bandwidth and energy efficiency.} On smartphones and tablets,
energy is a scarce resource and communication can be expensive. As a
consequence, tools that do not consider these factors are likely to be
abandoned by their users.  Almost all the techniques  designed and implemented
so far have been conceived with the assumption that the devices are connected
via wired links. Thus, they are not particularly efficient from this point of
view. 

In SmartProbe, the estimation of the bottleneck capacity has been designed to
be suitable for mobile devices, with bandwidth efficiency as a primary goal.
Reducing the amount of data transmitted and received, in turn, brings
advantages in terms of energy efficiency. Our techniques considerably reduce
the amount of traffic: up to 96\% for 3G cellular networks and up to 89\% for
Wi-Fi networks.

\noindent {\bf Tuning for wireless connections.} Most existing techniques do not
behave properly in wireless scenarios, which tend to have high bit error
rates.  In addition, the tools currently available for smartphones
implement rather trivial techniques: they perform an HTTP GET request and
calculate the time needed to retrieve the requested page. However what is
obtained is just an estimation of {\em TCP throughput} and not of bottleneck
capacity.  These are actually two different properties that should not be
confused~\cite{Jain:2003:EAB:941374.941378} (for instance, TCP throughput
depends on a number of factors such as buffer size, round trip time, and
retransmission errors).

SmartProbe introduces customizations to state-of-the-art techniques in order to
operate more smoothly in wireless scenarios. 

\noindent {\bf Support for possibly large numbers of users.} The use of
bottleneck estimation methods from a crowdsourcing perspective creates a new
set of problems. Countermeasures are needed to prevent simultaneous
measurements from interfering with each other. 

SmartProbe tackles these problems by incorporating scheduling and queuing
mechanisms (on the server-side) to support large numbers of
users.

\section{Estimation of bottleneck capacity: theory and state of the art}
\label{sec:theory}

\begin{figure}[t]
   \centering
   \includegraphics[width=1.0\columnwidth]{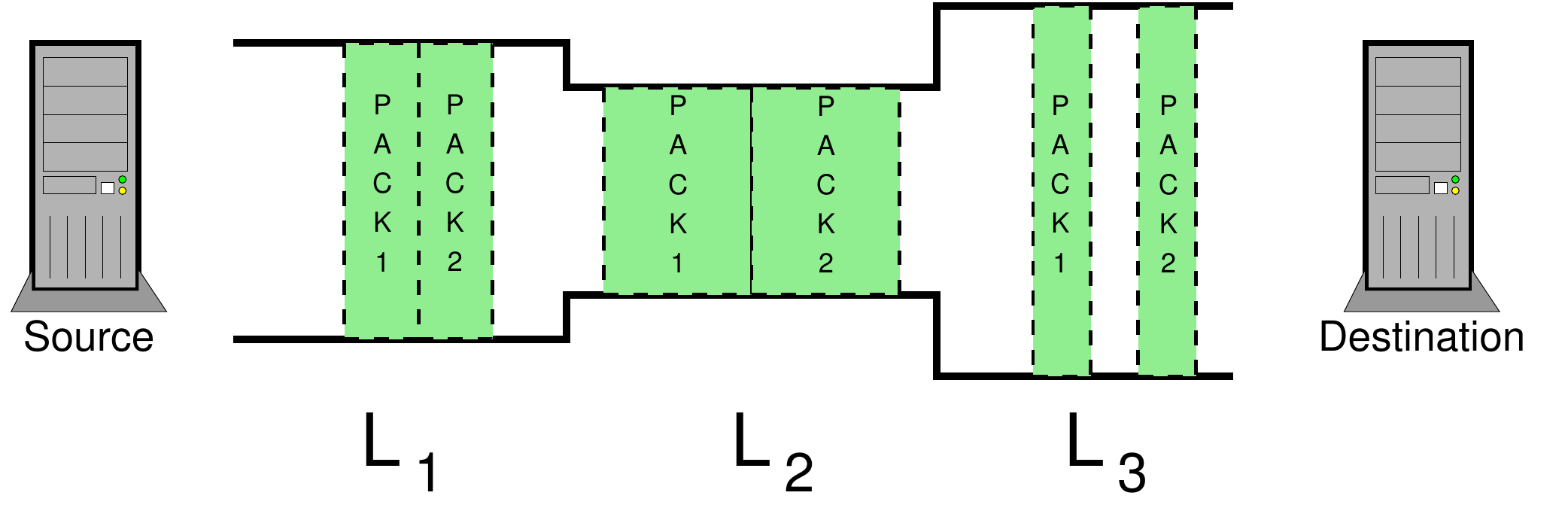}
   \caption{A pair of packets going through a set of links with different capacities.}
   \label{fig:packetpairs}
\end{figure}

Many capacity estimation techniques are based on the dispersion of a pair of
packets.  If link $i$ has capacity $C_i$, transmitting a packet with size $P$
on such link requires $\tau_i = P/C_i$. When two packets are sent back to back,
they reach the receiver with a dispersion $D = P/C$, where dispersion is the
time between the last bit of the first packet and the last bit of the second
packet.  If the transmission of the two packets entails going through $n$
links, the dispersion at the receiver will be $D = P/\min\{C_1, ..., C_n\}$.
Thus, by measuring on the receiver the dispersion of a couple of packets with a
known size, it is possible to calculate the bottleneck
capacity~\cite{Jain:2004:TFP:1028788.1028825,Dovrolis:2004:PTC:1046014.1046015}.
Figure~\ref{fig:packetpairs} shows a pair of packets (PACK1 and PACK2) going
through a set of links.  The smaller the capacity of a link, the longer the
time needed to transmit the packets. When the two packets reach the destination
host the dispersion depends on the capacity of $L_2$, which is the narrowest
link.  This technique forms the basis of a large set of tools including, for
instance, \textit{Pathrate}~\cite{Dovrolis} and
\textit{CapProbe}~\cite{CapProbe}).  In general, this procedure is repeated
several times (i.e. several packet pairs are sent) to obtain statistically
significant results.

Systems based on the dispersion of packet pairs may be inaccurate in the
presence of high speed links~\cite{Chen}.  Capacity is estimated as $C=P/D$,
but in high speed networks $C$ is large and $P$ is limited by the size of the
Maximum Transmission Unit (MTU). As a consequence, $D$ becomes very small,
which may lead to timer resolution problems. For example, let us consider a
1~Gbps network where MTU is 1500~B: in this case $D$ is equal to
$12 \mu s$, a value that cannot be easily measured with common system
timers. 

Another possible source of inaccuracy is interrupt coalescence, a technique in
which Network Interface Cards (NICs) generate a single interrupt for multiple
packets, when they are received in a short time interval~\cite{Prasad}.  With
interrupt coalescence, $D$ gets reduced because of data buffering and, in the
end, the resulting $C$ may be incorrect. 

These problems are solved by enlarging the numerator of the above introduced equation, i.e.
by sending a larger amount of data from source to destination. Since the MTU
cannot be enlarged indefinitely, the only possibility is to send a larger
number of packets.  The capacity can thus be computed as:\\

\begin{equation}
    C=\frac{(k-1)P}{D}
    \label{for:mula}
\end{equation}\\

where $k$ is the size of the train and $D$ is now the dispersion between the
first and last packet of the train.

\begin{figure}[t]
   \centering
   \includegraphics[width=1.0\columnwidth]{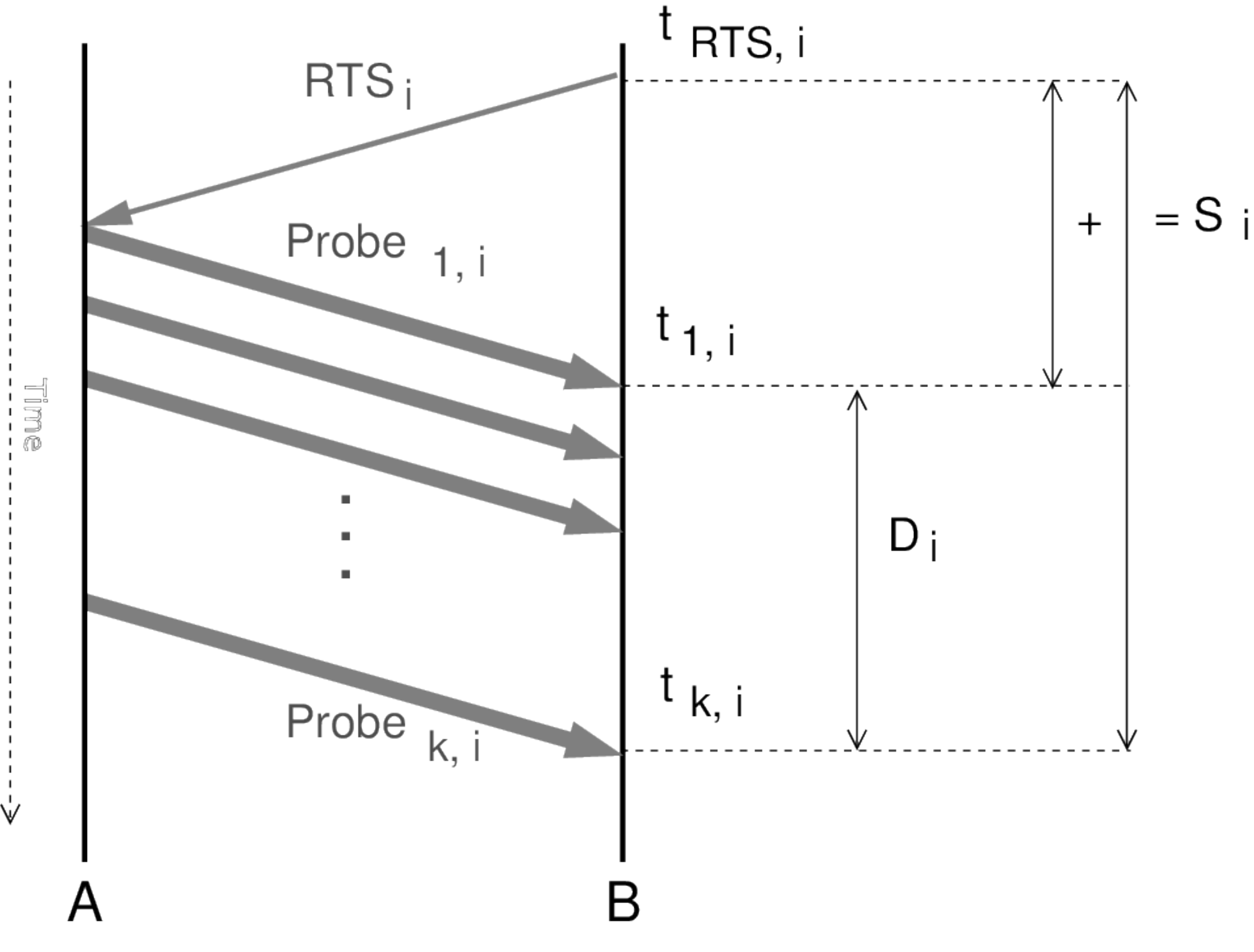}
   \caption{Delay sum components.}
   \label{fig:ds}
\end{figure}

The presence of cross-traffic in the path -- that is Internet traffic generated
by other sources and dispatched by the same router(s) -- can alter the
dispersion values. This may lead to capacity estimation errors. Moreover, since
packet trains are longer than packet pairs, techniques based on trains are more
prone to this problem than the ones based on pairs.

Some tools include mechanisms for reducing the effects of external noise.
PBProbe adopts the concept of \textit{delay sum} to minimize the under- and
over-estimations caused by cross-traffic. Figure~\ref{fig:ds} shows the
estimation procedure in PBProbe.  In particular, the
bottleneck capacity is estimated along the path that goes from A to B.  On
receiving a Request To Send (RTS), A sends a train of probes to B.  The delay
sum for the $i$-th train, indicated as $S_i$, is computed as the sum of the
delays experienced by the first and the last packet of that train. Thus, if $n$
trains are used, the minimum delay sum identifies the train that has
experienced the minimum queuing delay. Once the train in a campaign with the
minimum delay sum has been identified, the capacity is computed with
Equation~\ref{for:mula} using the dispersion $D$ experienced by such train.

PBProbe was originally designed to estimate the bottleneck capacity of
high-speed wired networks, but it was also successfully used to infer the
bottleneck capacity in wireless environments~\cite{Chen}. The consumption of
resources is the main issue that prevents the direct adoption of PBProbe on
mobile phones. To obtain a valid result, PBProbe uses $n=200$ trains, where
each train is composed of $k$ packets of 1500 bytes. The larger the value of
$k$, the larger the dispersion time $D$ experienced between the arrival of the
first and last packet of each train and, as a consequence, the smaller the
impact of potential interference phenomena caused by interrupt coalescence and
by timer resolution. To identify the correct value of $k$, PBProbe 
compares the dispersion time $D$ registered by trains with a
threshold value $D_{thresh}$: if any of the trains experience $D < D_{thresh}$,
then the train length is increased tenfold and the computation starts again
from scratch.  In~\cite{Chen}, $D_{thresh} = 1$~ms is considered the minimum value 
able to limit the impact of system interferences. This means that PBProbe
requires sending about two thousand packets, i.e. about 3~MB of data, to
correctly compute the capacity of a 802.11g wireless network. 

On devices with limited resources, such as smartphones, $D_{thresh} = 1$~ms is
not a realistic value.  As discussed in the following (in
Section~\ref{sec:selection}), the impact of interference phenomena on these
devices is acceptable when $D_{thresh} = 10$~ms. In this case, PBProbe would
require more than twenty thousand packets, i.e. about 30~MB of data, for a
802.11g network.  
In general, the amount of data produced by PBProbe to correctly infer the
bottleneck capacity of mobile networks (assuming that $D_{thresh}$ is equal to
10~ms) is between about 300~KB and 300~MB\footnote{300~KB is obtained when
considering the parameters of a slow connection like GPRS, a declining
technology. Common values are thus more likely in the higher end of the
range.}. This often represents a problem: a large amount of transferred data is
likely to lead to a significant consumption of energy and, depending on the
commercial agreement between users and mobile operators, it may also lead to
economic costs.

\section{Estimating the bottleneck capacity on smartphones}

This section presents an overview of the SmartProbe system and provides the
details of the bottleneck estimation procedure. 

\subsection{System overview}

\begin{figure}[t]
        \centering
        \includegraphics[width=1.0\columnwidth]{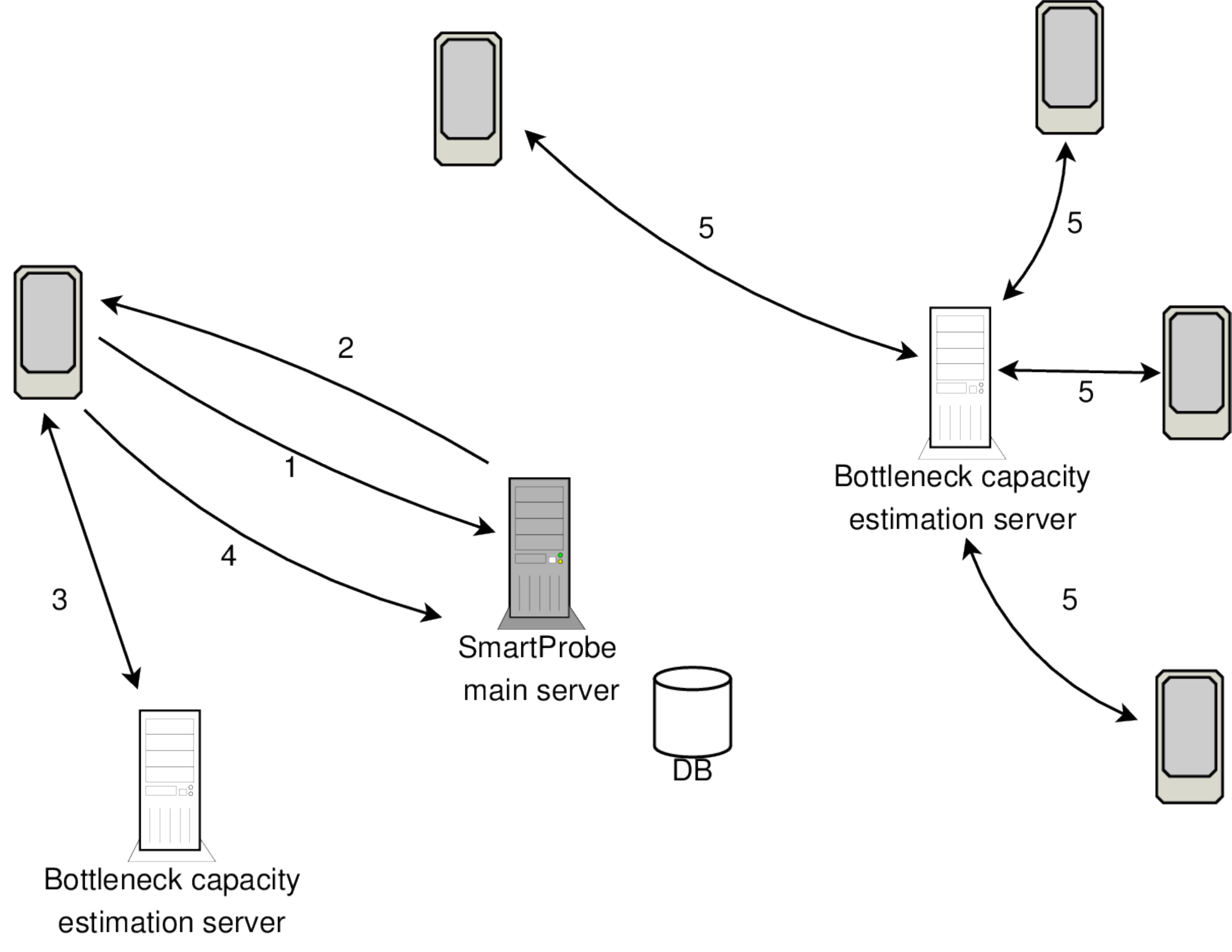}
        \caption{Overview of the system.}
        \label{fig:overview}
\end{figure}

As previously mentioned, measuring the bottleneck capacity along a path requires active
communication between the two endpoints. The source and destination hosts
exchange a set of packets, usually called probes, and collect timestamps to
infer this property of the network. The procedure is generally repeated 
several times to obtain accurate results. 

Figure~\ref{fig:overview} shows a high-level overview of the SmartProbe system.
A number of clients (smartphones) interact with the main server, which
coordinates their activities and acts as collector of results. Other
secondary servers are used as endpoints for bottleneck capacity estimation
measurements. Each individual bottleneck capacity estimation server (BCES) is
able to handle a number of clients.

When a user starts a measurement, the smartphone contacts the SmartProbe server
(1) which replies with the address of the BCES to be used as the measurement
endpoint (2). The selection of the best BCES can be based on the country of the
device or its geographical coordinates. The client then performs the required
operations by interacting with the given BCES (3). The result is the estimated
bottleneck capacity along the path that goes from the client to the selected
BCES. The measurement procedure is executed two times to obtain the bottleneck
capacity for the two directions. Results are then forwarded by the client to
the SmartProbe server (4), where they are saved onto persistent storage.

Measuring the bottleneck capacity does not require a large amount of time.
However, given that the system is designed to operate in a crowdsourcing
scenario, a single BCES has to cope with numerous users and it may be involved
in multiple measurements at the same time (as depicted on the right-hand side
of Figure~\ref{fig:overview}).  It is clear that without appropriate
strategies, simultaneous measurements (5) could be compromised or distorted
because of mutual interference.  For this reason, BCESs include mechanisms for
multiplexing and queuing measurement requests coming from different clients.
These methods are described in Section~\ref{sec:server}.

SmartProbe is currently available as a service provided by the Portolan
platform, a smartphone-based crowdsourcing system aimed at sensing 
large-scale networks. The SmartProbe functionalities, on the client-side, were
incorporated as a module of the Portolan app, which besides bottleneck
estimation provides other network tools (signal coverage maps and exploration
of the graph of the
Internet)~\cite{faggiani12:feasibility,gregori13:sensing,portolancommag}. The
Portolan app (for Android), and thus also SmartProbe, is available for free on
Google Play.

\subsection{Protocol overview}

\begin{figure}[t]
        \centering
        \includegraphics[width=0.8\columnwidth]{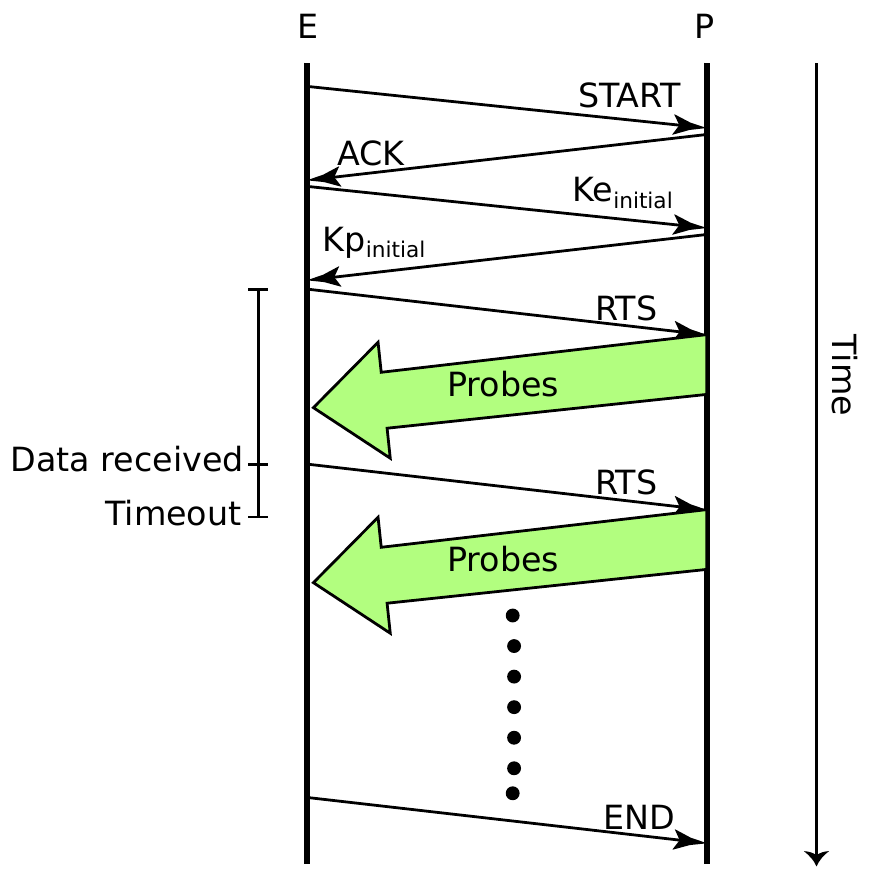}
        \caption{SmartProbe protocol.}
        \label{fig:protocol}
\end{figure}

The main steps of the bottleneck estimation procedure are shown in
Figure~\ref{fig:protocol}. Initially, one of the two hosts operates as
\textit{Estimator} (E) -- i.e.  the host that asks for packets and computes the
capacity -- while the other acts as \textit{Prober} (P), i.e. the host that
waits for requests and sends packet trains. Each phase begins with an UDP
handshake between Prober and Estimator to synchronize and initialize the two
hosts, followed by the transmission of the train of UDP packets.  The Estimator
starts the handshake by sending a START message to the Prober, which
replies with an ACK. After this, each host calculates the minimum $k$
that has to be used to compute the capacity without being affected by timer
resolution and interrupt coalescence problems (see Section~\ref{sec:pt}). The
smallest $k$ is chosen for the measurement, since the slower link introduces a
delay large enough to estimate the capacity correctly in both directions. 

The Estimator then starts the measurement by sending an UDP control message
named RTS (Request To Send), which triggers the dispatch of an UDP
train. The RTS message contains the number of packets $k$ that the
Prober has to use for the train it is going to send. The dispatch of every UDP
control message also starts a timer that allows the two hosts to not stall in
case of packet losses. Once the Estimator has correctly received all the
packets of a train, it computes the dispersion and then the capacity (using
Equation~\ref{for:mula}). Otherwise the train is invalidated when the timeout
expires. If multiple failures are experienced, the two hosts assume that the
network is congested and the experiment is restarted by halving the train
length. The estimation is completed when \textit{n} valid trains are received.
When this happens an END message is sent by the Estimator to the
Prober. Once finished, Prober and Estimator swap their roles and the procedure
is repeated, to obtain an estimation of the bottleneck capacity in the opposite
direction.

\begin{figure}[t]
\begin{lstlisting}
	@$i \gets 0$@
	@$failed \gets 0$@
	@$D_{min} \gets +\infty$@
	@$S_{min} \gets +\infty$@
	@$k \gets k_{initial}$@
	while(@$i < n$@)
		Set timeout
		Receive packet train @$i$@
		if (timeout triggered)
			@$failed \gets failed+1$@
			if (@$failed == 3$@)
				@$i \gets 0$@
				@$failed \gets 0$@
				@$k \gets max(\lfloor k/2 \rfloor , 2)$@
				@$D_{min} \gets +\infty$@
				@$S_{min} \gets +\infty$@
				continue
		else
			Measure @$D_{i}$@ @and@ @$S_{i}$@
			if (@$S_{i}$@ < @$S_{min}$@)
				@$S_{min} \gets S_{i}$@
				@$D_{min} \gets D_{i}$@
		@$i \gets i+1$@
	Compute capacity with @$D_{min}$@
\end{lstlisting}
\caption{SmartProbe estimation algorithm.}
\label{fig:alg_smt}
\end{figure}

The SmartProbe algorithm is expressed by the pseudo-code in
Figure~\ref{fig:alg_smt}.

\subsection{Dispersion and capacity}

\begin{figure}[t]
  \centering
  \includegraphics[width=0.5\columnwidth]{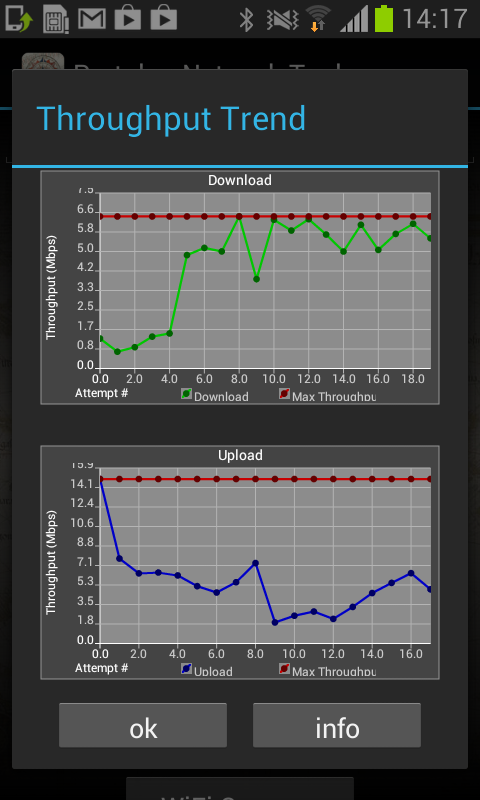}
  \caption{The capacity values obtained with the different trains are shown to the user.}
  \label{fig:tes}
\end{figure}

SmartProbe computes dispersion according to a procedure that enhances the
procedure based on delay sum.  Let us call $t_{1,i}$ the delay registered by
the first packet of the $i$-th train, and $t_{k,i}$ the delay of the last
packet of the same train. Let us also define ${t_1}^{min} = \min t_{1,j}, \quad
j \in 1..i$, i.e. the minimum interval registered so far between a request to
send and the arrival of the first packet of a train.  The capacity value for
the $i$-th train is then calculated using the dispersion value $D_i = t_{k,i} -
{t_1}^{min}$, and applying Equation~\ref{for:mula}.  A similar technique, for
packet pairs, has also been presented in~\cite{Chan:2009:MMM:1658939.1658963}.

This approach has been followed because the value of capacity for the $i$-th
train, calculated using $t_{k,i} - t_{1,i}$ as dispersion value, may generate an
over-estimation due to compression phenomena~\cite{Chen}.  In fact, a better
estimate of the minimum time needed to receive the first packet of a train may
be already available. In other words, we can state that if the first packet in
a train experienced an arrival time larger than the minimum, then this is due to
cross-traffic.  Therefore, a more reliable estimation of the capacity, for
the $i$-th attempt, is obtained when using $D_i = t_{k,i} - {t_1}^{min}$ as
the dispersion value. 

This procedure is repeated $n$ times (for all trains), then the capacity value
corresponding to the minimum delay sum is selected as the one that best
estimates the bottleneck.

Once the estimation is completed, the final results are shown to the user. Besides the
bottleneck capacity values in upload/download, SmartProbe displays all the values computed
during the estimation process. Figure~\ref{fig:tes} shows the measured capacity
against the train number.

\subsection{Selection of $D_{thresh}$ in a smartphone-based scenario}
\label{sec:selection}

In previous literature, $D_{thresh}$ is set to 1~ms on the basis of
experimental results obtained from wired servers. Unfortunately, this value
cannot be directly applied to a smartphone-based scenario, as these devices
are characterized by relatively scarce resources. Thus, we run a set of
experiments aimed at finding the smallest $D_{thresh}$ value that provides an
acceptable loss of accuracy (due to timer resolution and interrupt
coalescence). 

\begin{figure}[t]
        \centering
        \includegraphics[width=1.1\columnwidth]{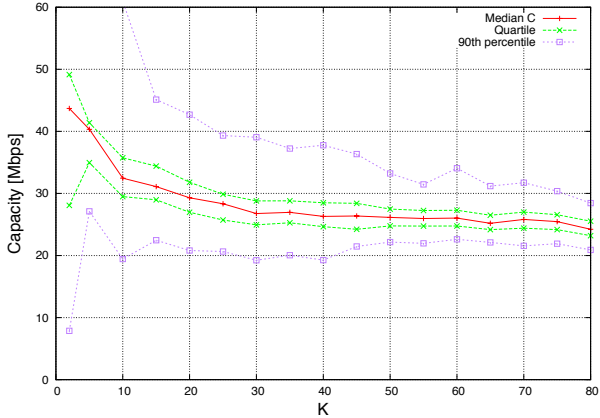}
        \caption{Stability of capacity estimation when varying the train length.}
        \label{fig:dispersion_quartiles}
\end{figure}

We followed an approach derived from the one described in~\cite{Chen}.
In particular, we analyzed the accuracy of smartphone-based measurements in an
unloaded Wi-Fi network. The experimental setup consisted of a smartphone
connected to an 802.11g wireless network and a BCES on the same wired LAN of
the wireless access point (Figure~\ref{fig:testbed}).  We varied the value of
$k$ to generate trains of increasing length (and thus characterized by
increasing dispersion). For each train length, the measurement has been
repeated 100 times. The results are depicted in
Figure~\ref{fig:dispersion_quartiles}: the average result of the tool converges
to a stable capacity value when using trains composed of more than 40 packets.
This value of $k$, by applying Equation~\ref{for:mula}, leads to a dispersion
value $D \sim 10$~ms, which we consider to be the minimum dispersion value
$D_{thresh}$ on smartphones (the effects of operating system interferences are
not negligible with smaller values).

\subsection{SmartProbe bandwidth-saving features}
\label{sec:pt}

SmartProbe includes bandwidth-saving features to make the estimation procedure
compatible with resource-constrained devices.

\subsubsection{Packet train length\label{sec:ptl}}

\begin{figure}[t]
        \centering
        \includegraphics[width=0.8\columnwidth]{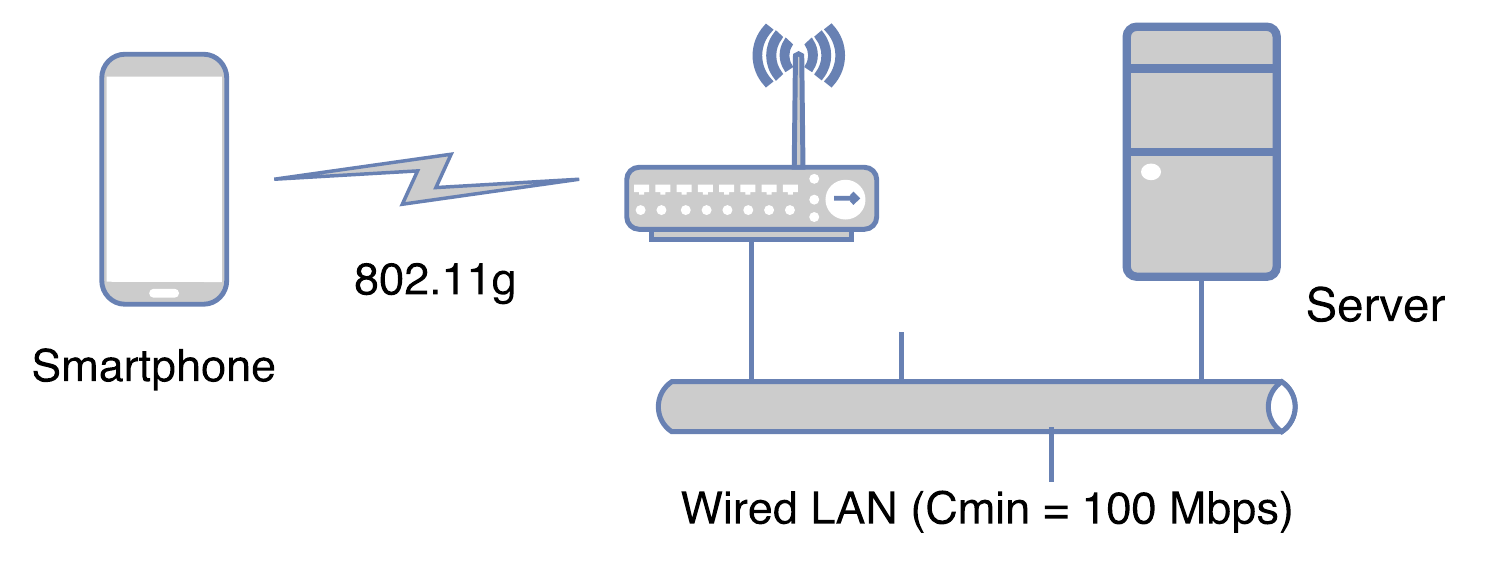}
        \caption{Experimental setup: the server is a Pentium 4,
2.80GHz machine equipped with 2 GB RAM running Linux Ubuntu 11.04,
whereas the smartphone is a Samsung Galaxy Nexus GT-19250 running Android
4.0.3; both the server and the access point are on the same
Ethernet LAN with 100Mbps NICs, while the smartphone is connected via
802.11g to the access point.}
        \label{fig:testbed}
\end{figure}

In PBProbe the value of $k$ is chosen dynamically by analyzing the dispersion
time: if a train is received with a dispersion value lower than a predefined threshold
$D_{thresh}=1$~ms, then $k$ is increased tenfold. 

SmartProbe follows a different approach: the a priori knowledge about the type
of network to which the smartphone is connected and the nominal capacity of
that type of network are used to determine the initial value of $k$
($k_{initial}$).  The type of network the smartphone is connected to is
directly provided by mobile operating systems. The nominal capacity typically
represents an upper bound of the real wireless performance~\cite{Cali}, thus it
can be used to understand the minimum number of packets required to produce a
dispersion value larger than $D_{thresh}$. The value of $k$ is dynamically
lowered by SmartProbe whenever too many trains experience packet losses.  This
can be caused for example by the presence of traffic shapers on the link or by
excessive contending traffic on the wireless access point. In these cases, the
application still tries to retrieve the value of the bottleneck capacity by
progressively lowering the value of $k$. The procedure stops when $k$ becomes
equal to 2, i.e. when a simple packet pair is used.

\begin{table}[t]
  \centering
\caption{Nominal capacity and initial train length per wireless technology.\label{tab:values}}{
	\begin{tabular}{lcrr}
		\hline
		\multicolumn{2}{c}{Network} & Cap. [Mbps]& $k_{initial}$ \\
		\hline
		\multirow{3}{*}{Wi-Fi 802.11} & \textit{b} & 11 & 11 \\
		 & \textit{a,g} & 54 &46 \\
		 & \textit{n} & 450 & 376 \\
		\hline
		\multirow{2}{*}{Mobile 2G} & \textit{GPRS} & 0.171 & 2 \\
		 & \textit{EDGE} & 0.473 & 2 \\
		\hline
		\multirow{2}{*}{Mobile 3G} & \textit{UMTS} & 1.8 & 3 \\
		 & \textit{HSPA} & 14.4 & 13 \\
		\hline
		Mobile 4G & \textit{LTE} & 326.4 & 273 \\
		\hline
	\end{tabular}	
}
\end{table}

The initial value of $k$, from which the computation starts, is obtained by
applying Equation~\ref{for:mula}, where $C$ is the nominal capacity of the
network type currently in use, $P$ is 1500 B and $D$ is $D_{thresh}$.
Table~\ref{tab:values} summarizes the values of $k_{initial}$ for common
wireless networks: the faster the network, the larger $k_{initial}$.

\subsubsection{Number of trains per campaign}

\begin{figure}[t]
        \centering
        \includegraphics[width=1.1\columnwidth]{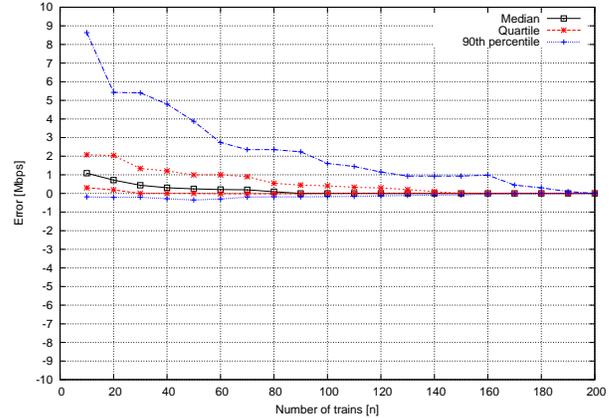}
        \caption{Error when varying the number of trains (with respect to 200 trains).}
        \label{fig:trains}
\end{figure}

Another factor that makes previous techniques less usable on smartphones is
that they use a fixed amount of trains for each measurement (e.g. 200 trains,
as previously mentioned). To decrease the amount of traffic -- and therefore
the costs in terms of connection fees and battery consumption -- \textit{n}
needs to be significantly reduced. We thus performed a campaign of 100
experiments with $n=200$.  In each test we first retrieved the most reliable
capacity $C^{200}$ using the full sample list. Then, we computed the capacity
$C^{m}$ that would be obtained by considering only the first $m$ trains.
Finally, we calculated the error between $C^{200}$ and $C^{m}$.
Figure~\ref{fig:trains} shows the distribution of errors when varying $m$
(median value, inter-quartile and 90th percentile). The median value of the
error becomes close to zero for $m\ge60$, meaning that $n=m=60$ is sufficient
to infer the correct capacity value most of the times.  Obviously, reducing the
number of trains leads to an unavoidable loss of accuracy. However in an
environment where saving energy is fundamental, the choice of $n=60$ represents
a good trade-off between accuracy and performance.

\section{Supporting a large number of users}
\label{sec:server}

In a crowdsensing scenario, it is likely that multiple users will want to
measure their bottleneck capacity simultaneously.  This requires the presence
of proper mechanisms on the server side, so that multiple estimations can be
carried out without interfering with each other.

\subsection{SmartProbe infrastructure: server side}
\label{sec:ss}

\begin{figure}[t]
  \centering
  \includegraphics[width=0.8\columnwidth]{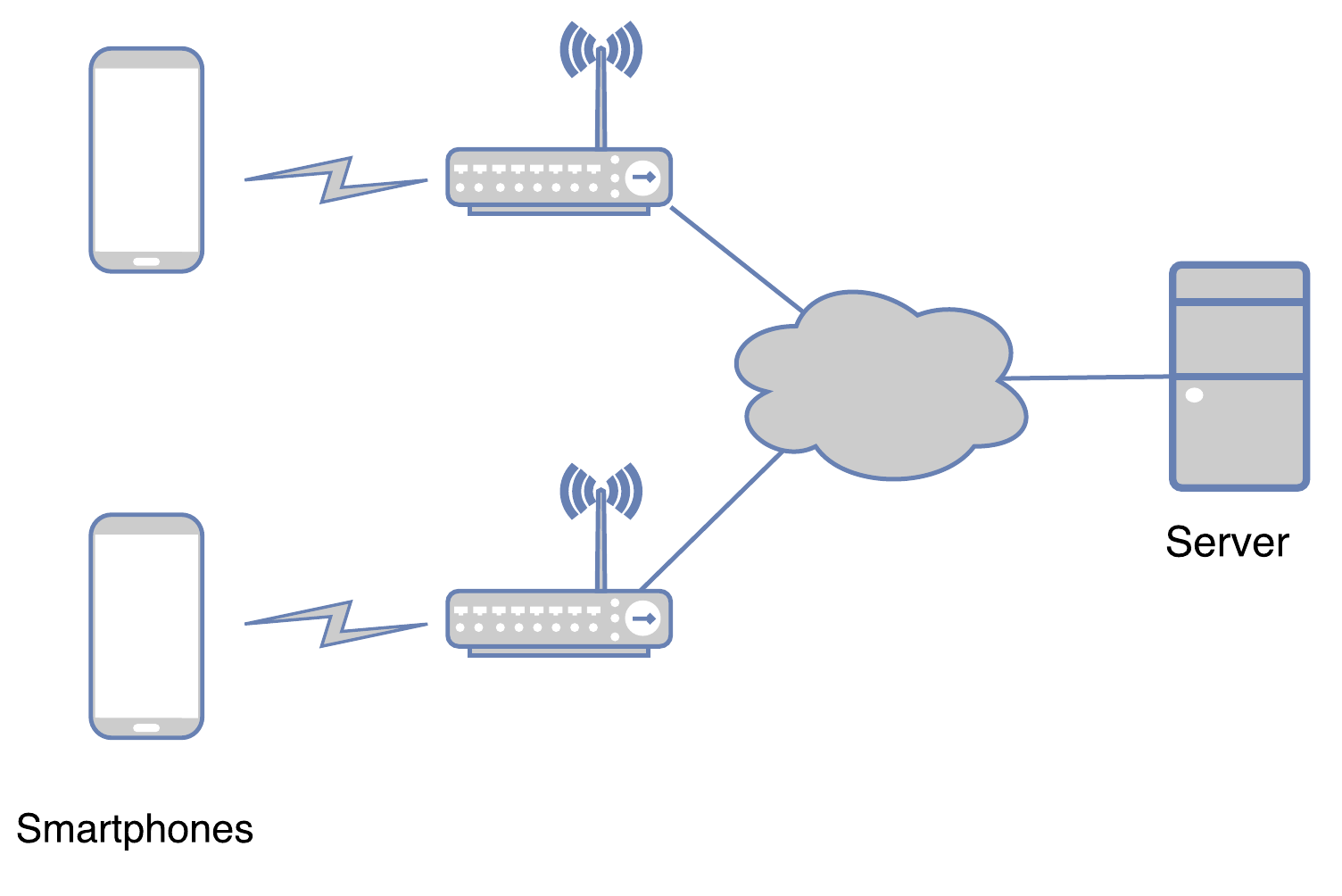}
  \caption{Bottleneck capacity estimation servers are connected to the Internet through a single link.}
  \label{fig:lasthop}
\end{figure}

The main problem in satisfying several requests at the same time is that the
computation and traffic load generated by multiple clients may reduce the
accuracy of measurements. In fact, requests from different sources converge to
the same BCES using a single link (Figure~\ref{fig:lasthop}).  Similar problems
may occur when multiple clients act as Estimators. For instance, suppose that
one client sends an RTS to the server; if the server is busy because
it is satisfying another request, it will not be able to respond immediately to
the new client.

\begin{figure*}[t]
  \centering
  \includegraphics[width=1.9\columnwidth]{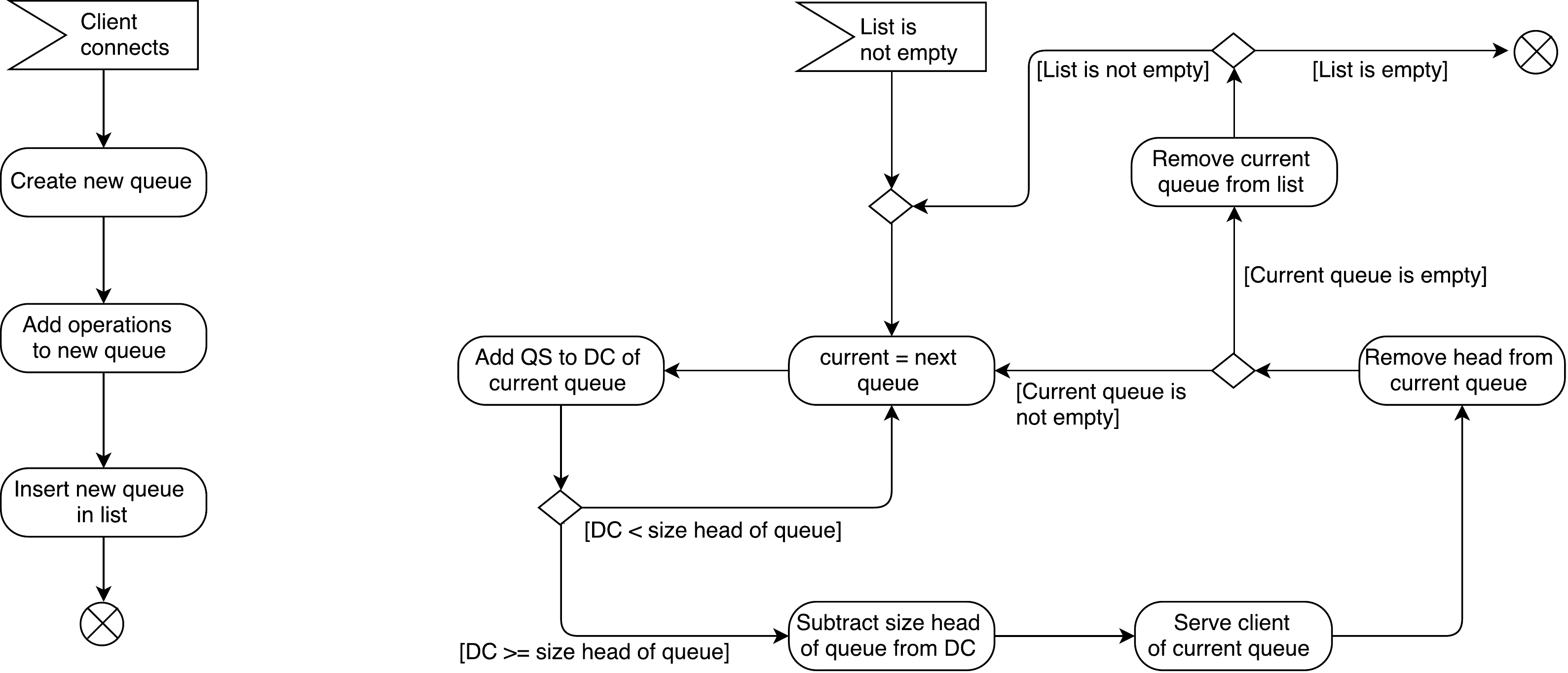}
  \caption{Activity diagram of scheduling}
  \label{fig:sm}
\end{figure*}

These problems can be avoided by preventing multiple clients from entering the
active measurement phase at the same time. To this purpose, the server includes
mechanisms for scheduling operations with clients.  The execution time of a
client action (i.e. sending a train of packets) depends both on its
geographical position, which influences the RTT, and the connectivity type, as
the length of the train changes with the type of network (Wi-Fi, UMTS, etc.).
As a consequence, slower (or more distant) clients may keep the server busy
more than faster (or closer) ones.  Thus, each BCES adopts a Deficit Round
Robin (DRR) scheduler for serving the requests coming from
clients~\cite{shreedhar96:drr}.

\begin{figure*}[t]
        \centering
        \subfloat[Deficit round robin: QS is added to DC of the first queue.]{\label{fig:Drrsub1}\includegraphics[width=0.9\columnwidth]{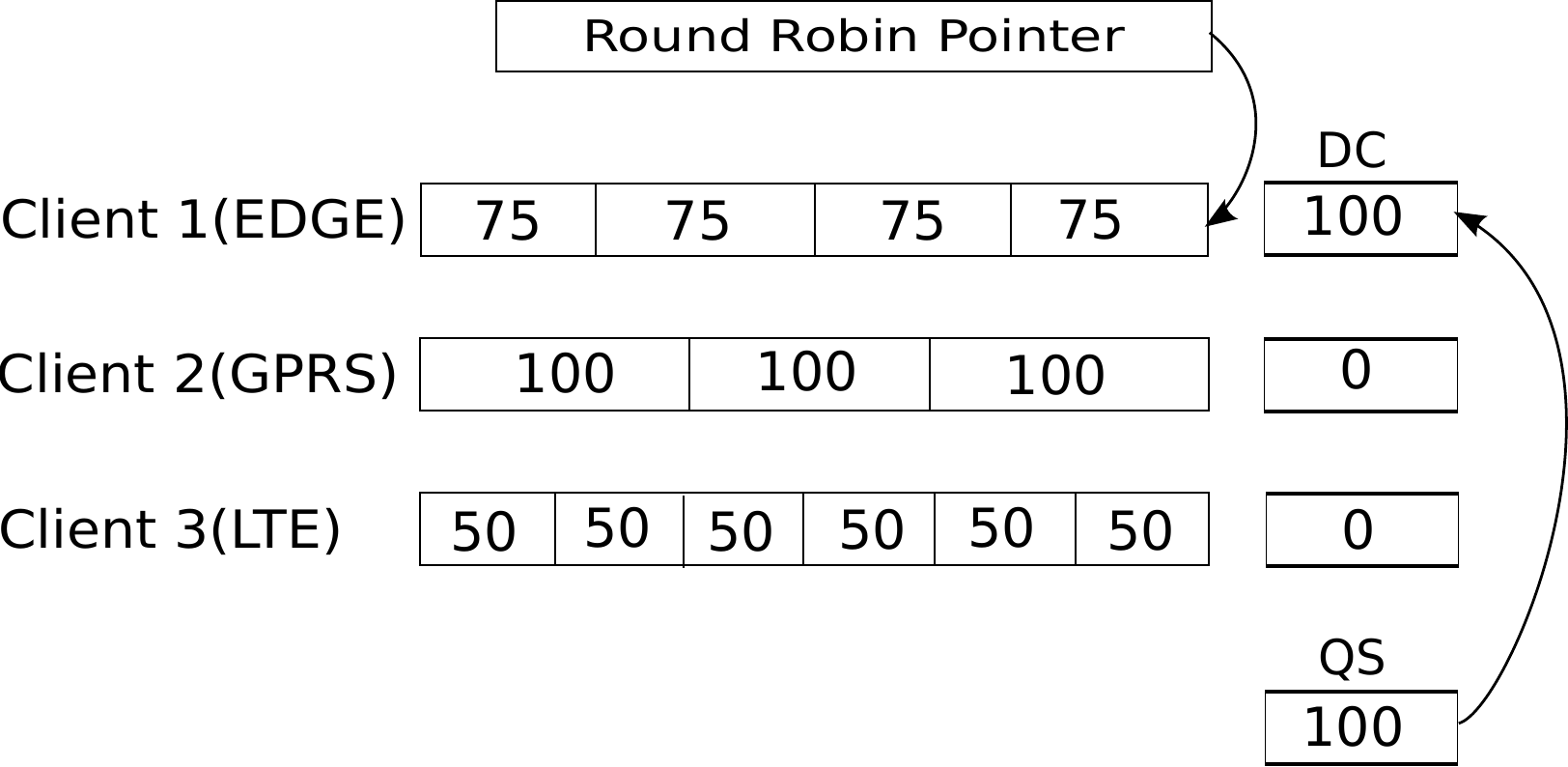}}
        \hspace{1cm}
        \subfloat[Deficit round robin: the first element of the first queue is scheduled, QS is added to DC of the second queue.]{\label{fig:Drrsub2}\includegraphics[width=0.9\columnwidth]{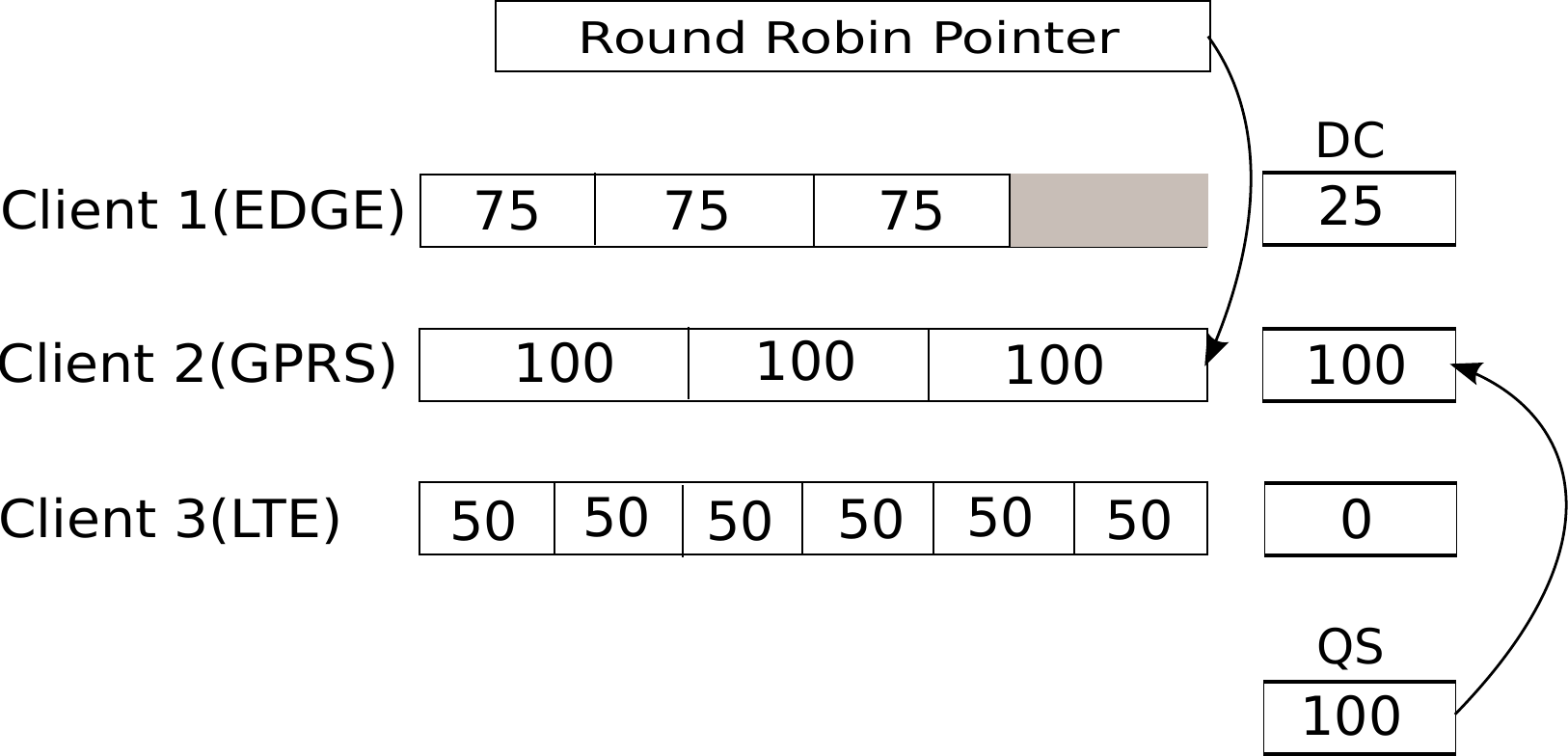}}
        \caption{Client operations are scheduled according to a Deficit Round Robin policy.}\label{fig:Drr}
\end{figure*}

Each requesting client is represented as an independent flow of the DRR scheme,
where each flow has its own queue (as shown in Figure~\ref{fig:Drr}).  In other
words, the list of flows/queues represents the set of clients currently involved
in an estimation. When an estimation is completed, the associated
queue is removed from the list. The size of the elements entering each DRR
queue (using DRR terminology) corresponds to the time a given client is assumed
to need to perform its next actions.
Let $V_{net,i}$ be the time needed by client $i$, connected via network type
$net$, to perform its next measurement. $V_{net,i}$ can be calculated as the
RTT between the client considered and the server, plus the time it takes to
send a train composed of $k$ packets, each $P$ bytes long, over a network with
nominal capacity $C_{nom}$ (i.e., $V_{net,i} = RTT_{net} + (k_{net,i} *
P)/C_{nom}$).

A Deficit Counter (DC) is associated with every queue.  The scheduling
algorithm is described by the diagram shown in Figure~\ref{fig:sm}: the two
depicted activities may occur in parallel and are event-triggered. When the
server receives a connection request from a new client, a new queue is created
and measurement operations are added to that queue (on the left-hand size of
Figure~\ref{fig:sm}). The size of operations is calculated as mentioned above.
The new queue is then inserted in the list of currently managed clients.  As
soon as the list becomes non empty, the activity depicted in the right-hand
side of Figure~\ref{fig:sm} is triggered. One of the queues is selected as the
current one, and an amount equal to Quantum Size (QS) is added to the DC of
that queue.  If the value contained in DC is greater than the size of the head
element of the current queue, the measurement operation can be executed,
otherwise the next queue is processed. If the operation is executed, the amount
corresponding to its size is removed from DC and the client is served. The just
processed operation is removed from the queue. When a queue becomes empty it is
removed from the list. If the list is not empty the next queue is processed,
otherwise the server goes idle and waits for new connections from clients.

Figures~\ref{fig:Drrsub1} and~\ref{fig:Drrsub2} show a scenario where three
clients are simultaneously interacting with the server. The three clients are
connected to the Internet using different access technologies: EDGE, GPRS, and
LTE. The duration of the probing operations is assumed to be equal to 100 with
GSM, 75 with EDGE, and 50 with LTE. In Figure~\ref{fig:Drrsub1} a QS equal to
100 is added to the DC of the first queue, then the first element of this queue
can be processed as the value of DC is greater than 75. The remaining amount
(25) is left in DC for the next round (Figure~\ref{fig:Drrsub2}). The round
robin pointer moves to the second queue and QS is added to its DC. Also in this
case the value in DC is sufficient for processing the first element of the
second queue (but in this case 0 will be left in DC). This procedure is
repeated until all clients have completed their operations. 

\begin{figure}[t]
  \centering
  \includegraphics[width=1.0\columnwidth]{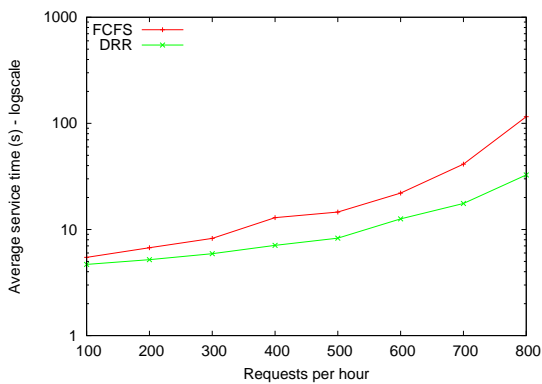}
  \caption{Service time with DRR and FCFS.}
  \label{fig:drr-fcfs}
\end{figure}

We studied the behavior of our DRR-based solution with respect to a system
based on a First Come First Served (FCFS) policy. Evaluation was carried
out through numerical analysis. The number of measurement requests per hour
was varied between 100 and 800, to study the behavior of the two policies under
different workloads. Communication between clients and server was characterized
by RTTs uniformly distributed between 0 and 20~ms. In the considered scenario, all
access technologies were equally represented.  Figure~\ref{fig:drr-fcfs} shows
the obtained results in terms of service time (the amount of time between the
time a request is issued by a client and the time the estimation is
completed). As expected DRR and FCFS operate similarly when the the load is
light: as soon as a request is received it can be immediately served,
independently from the scheduling algorithm. On the contrary, when the load is
high the differences between the two scheduling policies become significant. In
particular, for the highest request rate, the service time with DRR is $\sim70\%$
smaller than the one obtained with FCFS. 

To have a DRR \emph{WorkQuotient} equal to O(1), QS should be equal to the
maximum time a client may need to perform an action (as specified
in~\cite{shreedhar96:drr}). In our scenario, this could be done by using the
parameters of the slowest (in terms of capacity value) connection the system
has to cope with ($QS=RTT_{slowest} + (k_{slowest}*P)/C_{slowest}$), in
practice by using the parameters of the GPRS connection.  Nevertheless, we
observed that selecting the value of QS according to this principle provides
less benefits with respect to the ones shown in Figure~\ref{fig:drr-fcfs},
which have been obtained using a smaller value (approximately 1/3). This is
explained by the fact that in our scenario flows are not backlogged, and thus
using a smaller QS provides increased fairness.

\section{Validation and experimental results}

\begin{figure*}
        \centering
        \subfloat[Wi-Fi - download]{\label{fig:wifi-down}\includegraphics[width=0.8\columnwidth]{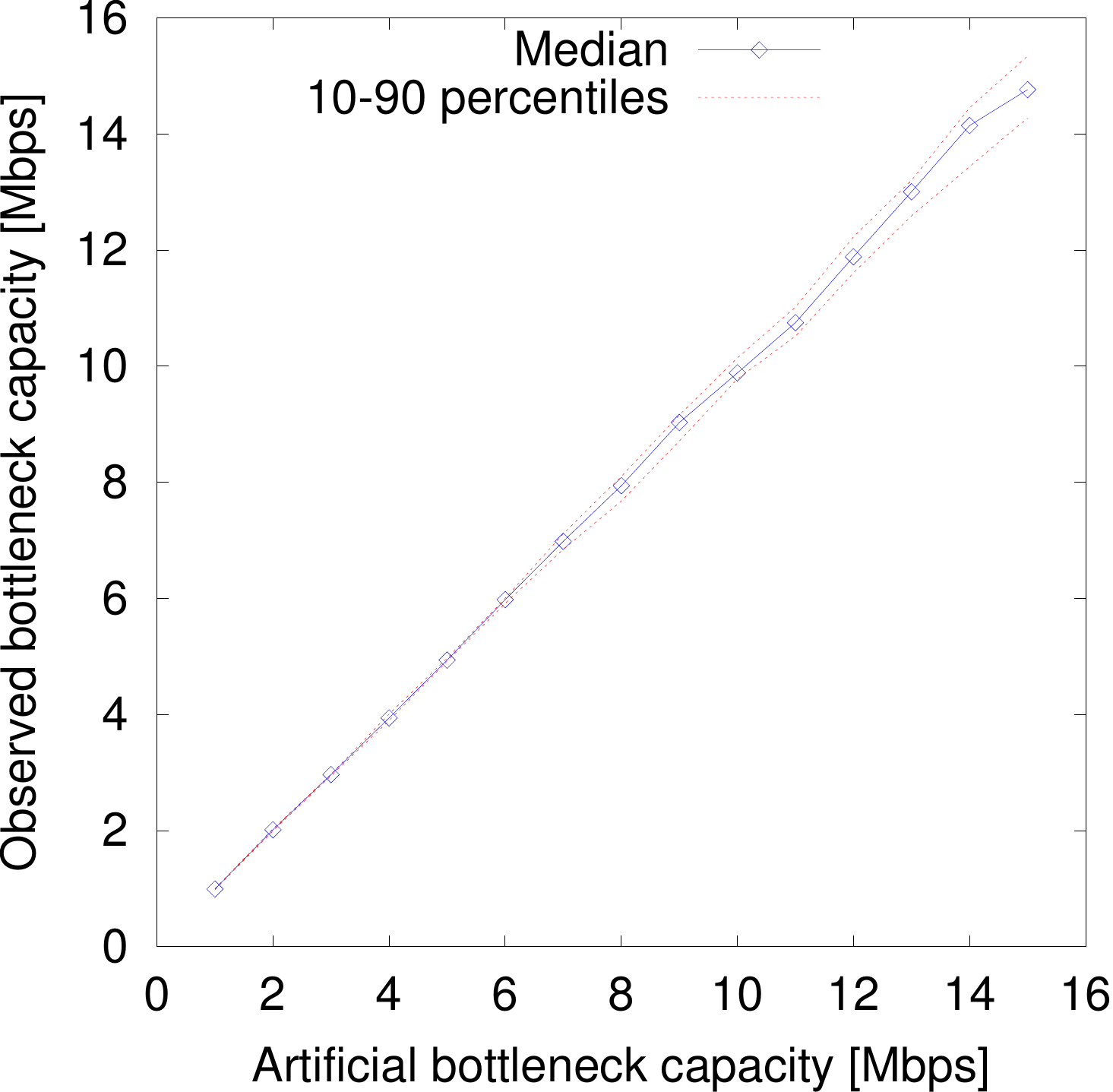}}
        \quad
        \subfloat[Wi-Fi - upload]{\label{fig:wifi-up}\includegraphics[width=0.8\columnwidth]{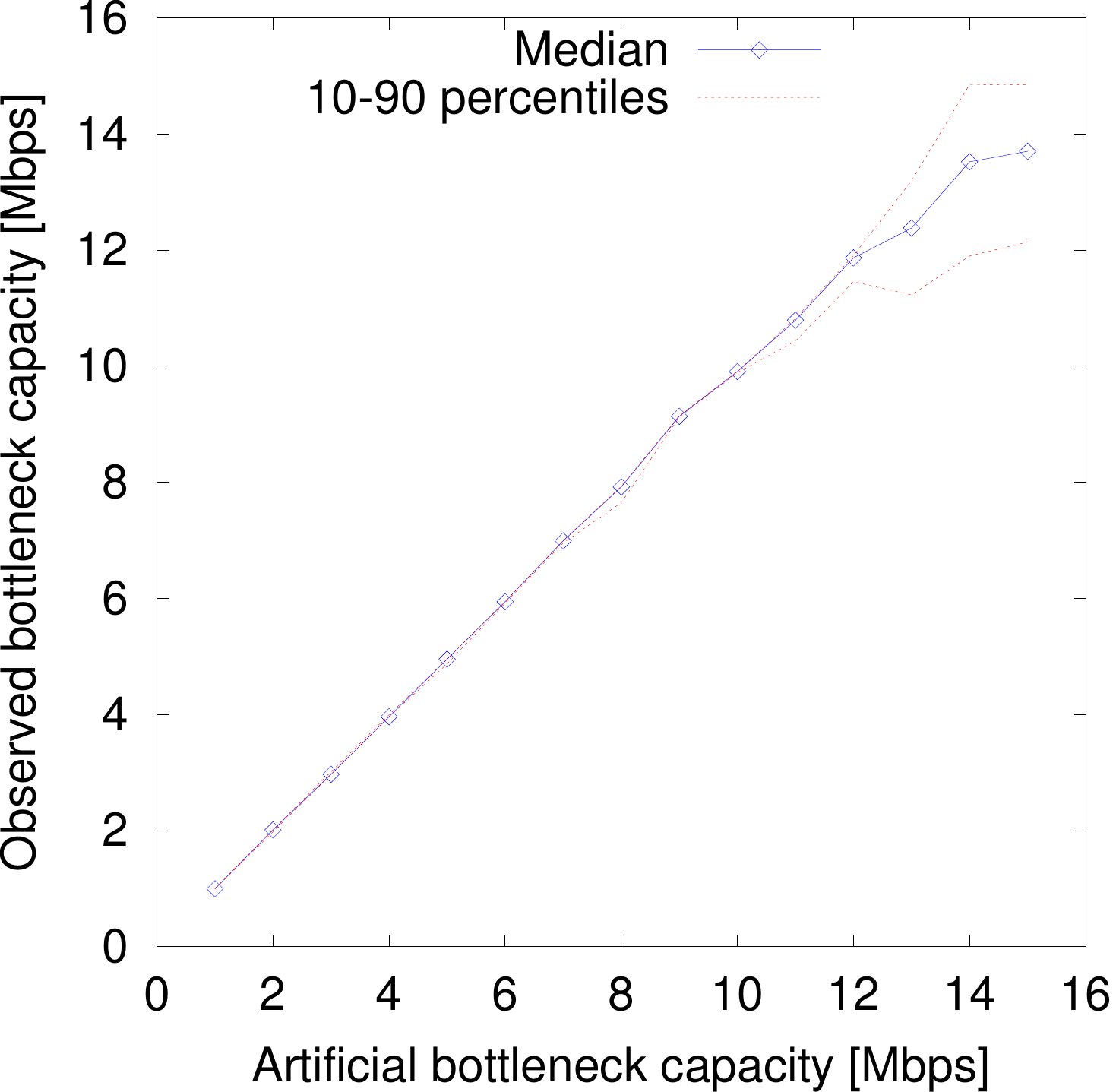}}
        \\
        \vspace{1cm}
        \subfloat[Cellular - download]{\label{fig:cellular-down}\includegraphics[width=0.8\columnwidth]{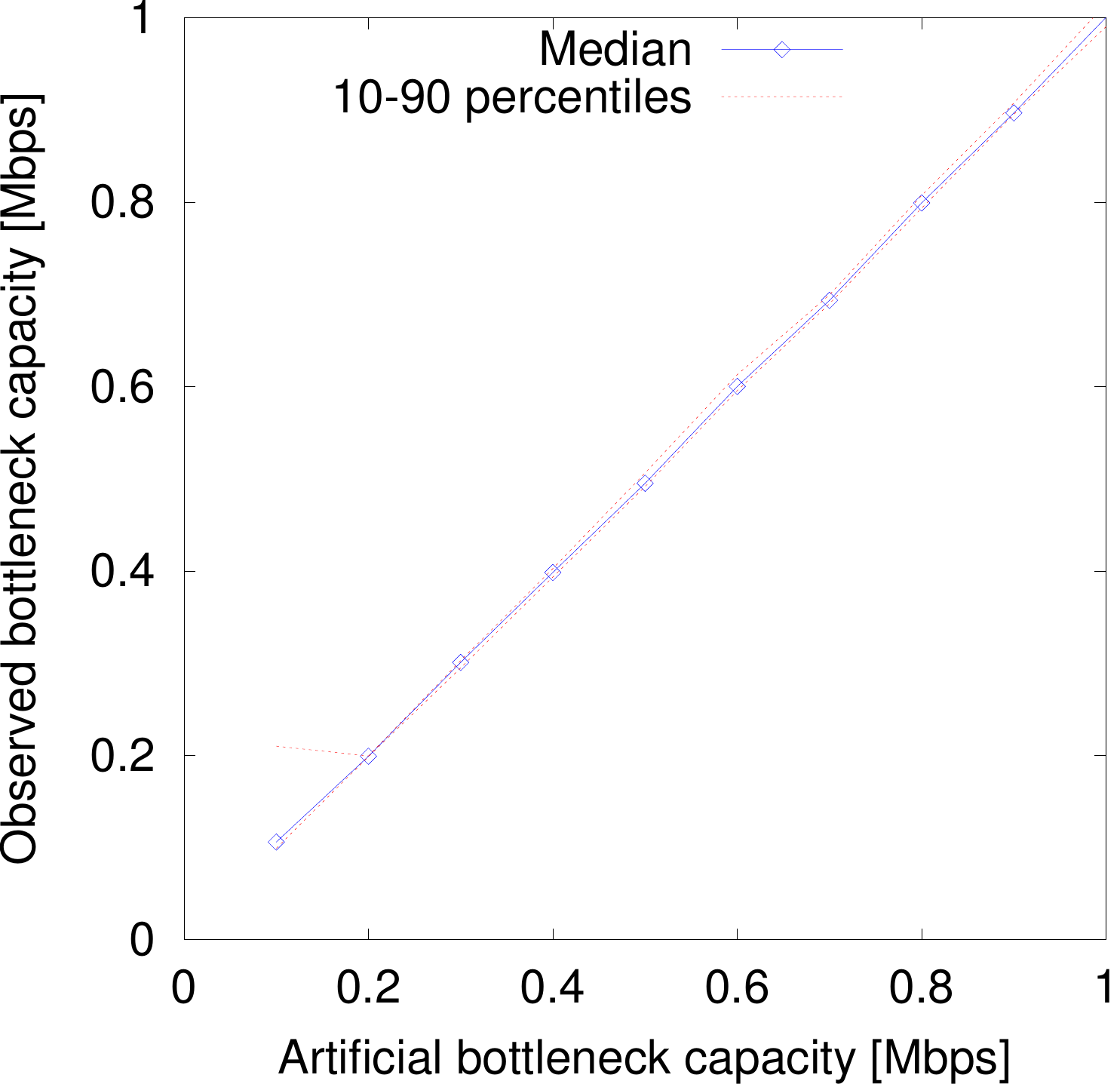}}
        \quad
        \subfloat[Cellular - upload]{\label{fig:cellular-up}\includegraphics[width=0.8\columnwidth]{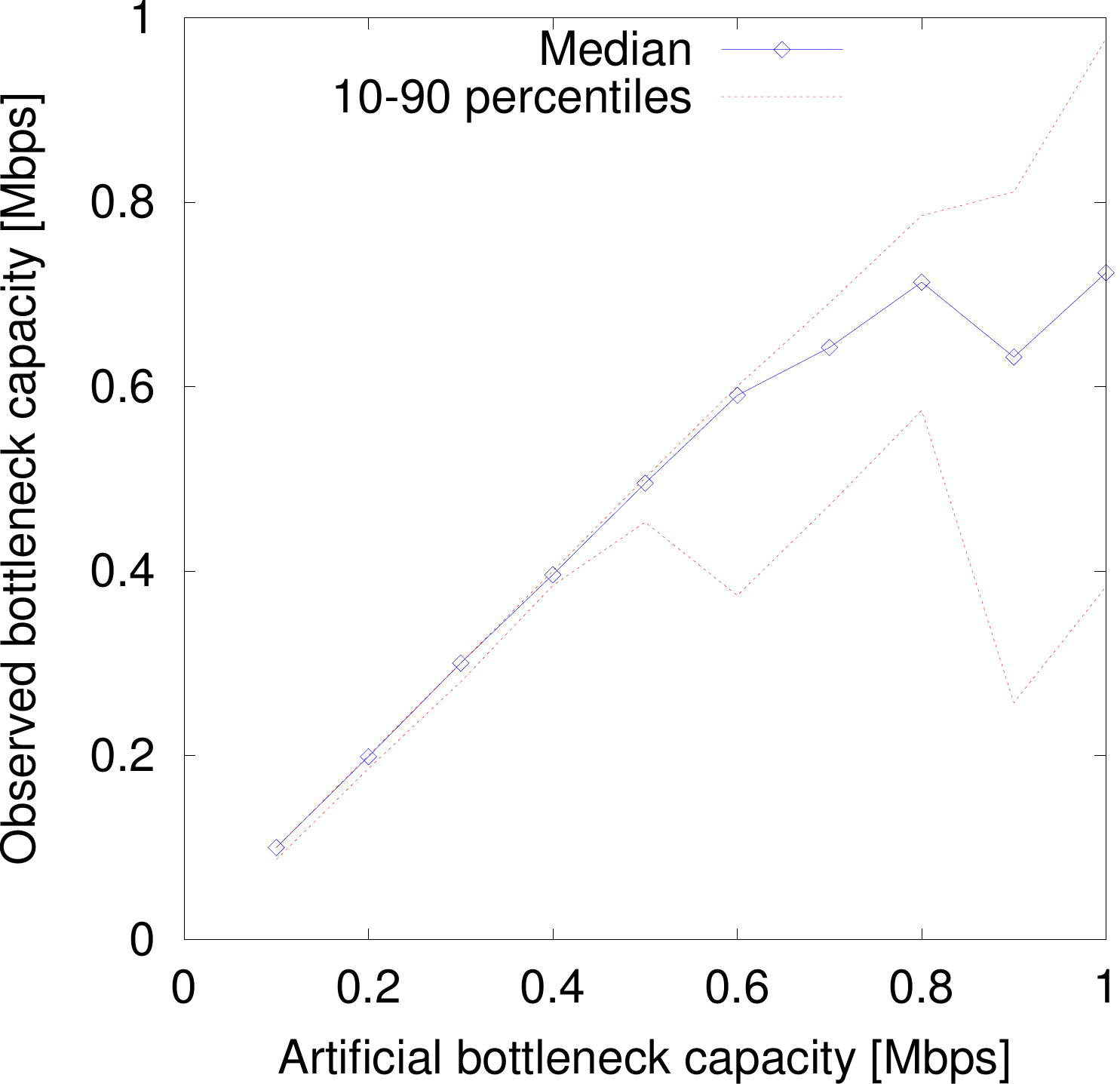}}
        \caption{Validation: observed bottleneck capacity against bandwidth limitation imposed on the server.}\label{fig:validation}
\end{figure*}

The mechanisms behind SmartProbe and their implementation were validated by
measuring the bottleneck capacity in a scenario where the ground truth was
known in advance. The scenario consisted of a smartphone running the SmartProbe
client and a BCES connected to the Internet (Figure~\ref{fig:testbed}). The
bandwidth of the connection between the BCES and the Internet was artificially
limited to a set of known values using Dummynet~\cite{carbone10:dummynet}. In
the first configuration, smartphone connectivity was obtained via Wi-Fi through
a local access point. Results are shown in Figures~\ref{fig:wifi-down}
and~\ref{fig:wifi-up}: for the considered range (1~Mbps - 15~Mbps) all the
capacity values registered by SmartProbe perfectly correspond to the values set
using Dummynet (for both directions).  In the second configuration, the
smartphone was connected to the Internet via its cellular interface.  The
considered range was 0.1~Mbps - 1~Mbps and the obtained results are shown in
Figures~\ref{fig:cellular-down} and~\ref{fig:cellular-up}.  Again, in download,
the capacity value measured using SmartProbe exactly matches the bandwidth
limitation imposed by Dummynet. In contrast, as far as upload is concerned, the
measured value and the imposed value start to diverge when the artificial
bandwidth limitation becomes greater than approximately 600~Kbps.  This is due
to the fact that SmartProbe measures the bottleneck capacity of the whole path
between the smartphone and the BCES involved. In fact, when the limitation
imposed by Dummynet is large, the bottleneck capacity is not located on the
link that connects the server to the Internet, but on the cellular connection
between the smartphone and the base station.  Thus we can reasonably say that
in all the considered cases SmartProbe was able to provide an exact measure of
the bottleneck capacity (when the bottleneck is the artificial one on the link
between BCES and the Internet) or a credible value (when the bottleneck is
located on the uplink between the smartphone and the cellular base station).

\begin{figure}[t]
  \centering
  \includegraphics[width=1.0\columnwidth]{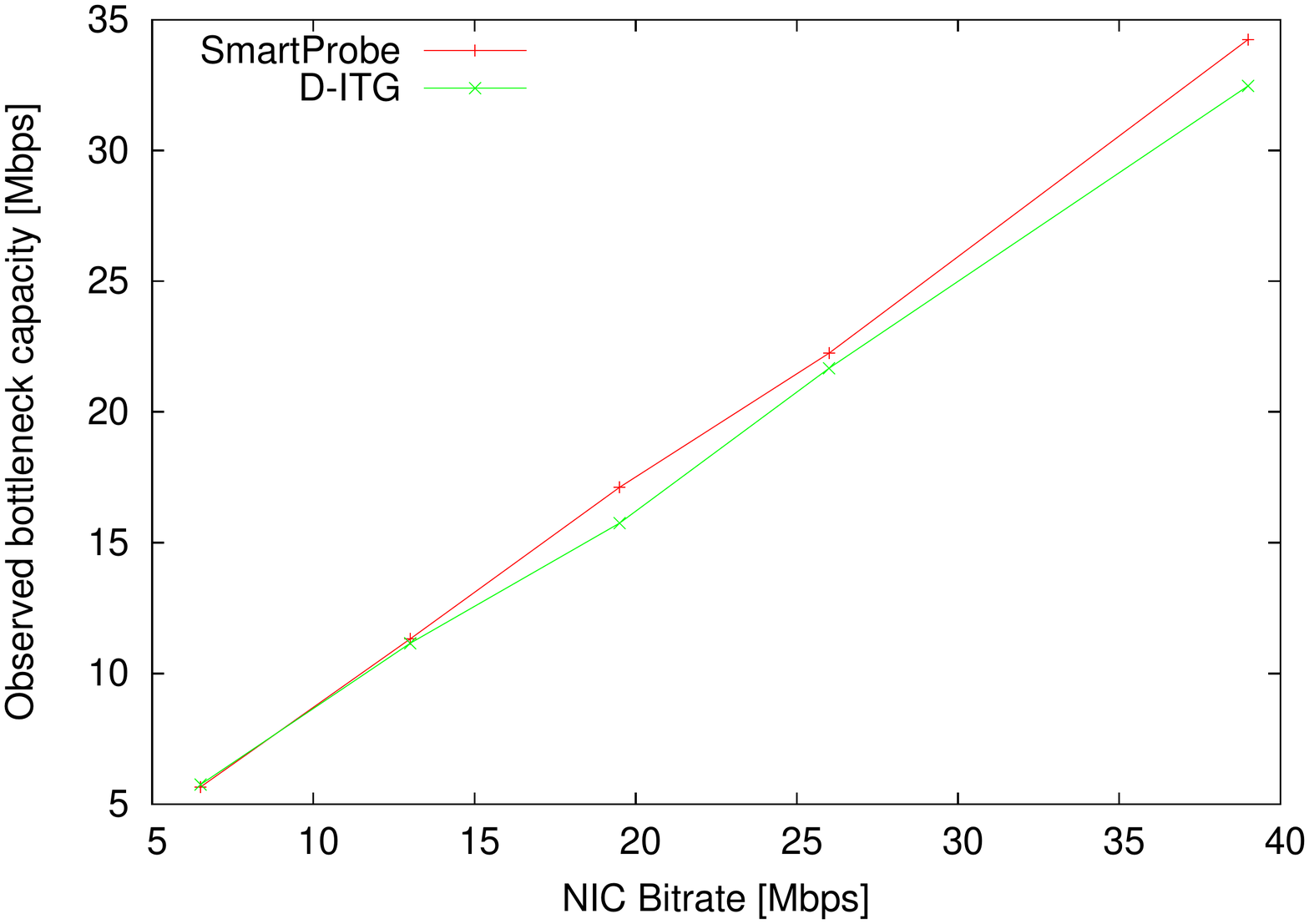}
  \caption{Validation when the bottleneck is located on the wireless link: capacities reported by SmartProbe and D-ITG.}
  \label{fig:smart-ditg}
\end{figure}

To further validate SmartProbe, the system was also tested in a configuration
where the link with narrowest capacity was the wireless one.  This additional
validation was carried out because in many real situations the wireless link
acts as the bottleneck. The considered scenario is similar to the one depicted in
Figure~\ref{fig:testbed}. The capacity of the wireless link was artificially
restricted by forcing the bitrate of the wireless interface to specific values
(the same approach has been used in~\cite{Sun}). The values we used correspond
to five of the slowest supported bitrates of IEEE 802.11n. The capacity
estimated by SmartProbe was compared to the one reported by
D-ITG~\cite{DBLP:journals/cn/BottaDP12}, which is not only a popular and
reliable platform for generating traffic, but also a network measurement tool.
Figure~\ref{fig:smart-ditg} shows the results averaged over ten repetitions
(only download for the sake of brevity, same findings have been obtained for
upload): the capacities reported by SmartProbe and D-ITG are very similar for
all the considered modes of operation. The average absolute difference between
the two tools is $\sim3.9\%$. We believe this is a rather small value for this
type of measurements, especially considering that the standard deviation of
SmartProbe results is, on average, equal to $\sim 1.3\%$ of the reported
capacity.

\begin{table*}[t]
\centering
\caption{SmartProbe: traffic generated in comparison with PBProbe\label{tab:comp}.} {
  \begin{tabular}{lcrrr}
    \hline
    \multicolumn{2}{c}{Network} & PBProbe [MB] & SmartProbe [MB] & SmartProbe worst case [MB]\\
    \hline
    \multirow{3}{*}{Wi-Fi 802.11} & \textit{b} & 3.3 & 1.0 & 1.6 \\
     & \textit{a,g} & 30.3 & 4.1 & 7.8 \\
     & \textit{n} & 300.5 & 33.8 & 67.1 \\
    \hline
    \multirow{2}{*}{Mobile 2G} & \textit{GPRS} & 0.6 & 0.2 & 0.2 \\
     & \textit{EDGE} & 0.6 & 0.2 & 0.2 \\
    \hline
    \multirow{2}{*}{Mobile 3G} & \textit{UMTS} & 3.3 & 0.3 & 0.4 \\
     & \textit{HSPA} & 30.3 & 1.2 & 2.2 \\
    \hline
    Mobile 4G & \textit{LTE} & 300.5 & 24.6 & 48.8 \\
    \hline
  \end{tabular}
}
\end{table*}

We evaluated the bandwidth-saving characteristics of SmartProbe. To this
purpose, we compared the amount of traffic generated by SmartProbe with the one
produced by a technique based on the PBProbe algorithm. We considered
$D_{thresh}=10$~ms for our smartphone-based scenario, as previously discussed. 

Let us suppose that the bottleneck capacity is imposed by a 802.11g link.  The
technique based on PBProbe initially sends a train composed of two packets. The
dispersion registered by such packet pair is below the threshold, thus the
length of the train (excluding the first packet) is increased tenfold, and
another train composed of eleven packets is sent. Also this train experiences a
dispersion that is below the threshold, thus the length of the train is again
increased tenfold. The train length is now sufficient for obtaining $D \ge
D_{thresh}$ and other trains with the same length are sent until $n=200$ trains
are received. This approximately generates 30.3~MB of data. SmartProbe starts
the estimation using $n=60$ and $k=46$ (see Table~\ref{tab:values}) for a total
of approximately 4.1~MB, about 13\% of the traffic generated by PBProbe. In
addition, even considering the worst case in which a persistent packet loss
causes the length of packet trains to decrease to the smallest value, i.e.
$k=2$, SmartProbe would require 7.8~MB of data\footnote{In the worst case,
SmartProbe would send firstly $n=60$ trains with $k=46$, then $n=60$ trains
with $k=23$, and so on with $k=11,5,2$, leading to a total of 5220 packets,
i.e. 7.8~MB of data}, which is still 26\% of data sent by PBProbe. Results for
other types of wireless networks can be found in Table~\ref{tab:comp}. The
amount of traffic is reduced up to 96\%. By sending a smaller amount of data,
SmartProbe obtains obvious benefits in terms of battery consumption and
economic costs.

\begin{figure*}[t]
        \centering
        \subfloat[Wi-Fi - download]{\label{fig:wifi-back-down}\includegraphics[width=0.95\columnwidth]{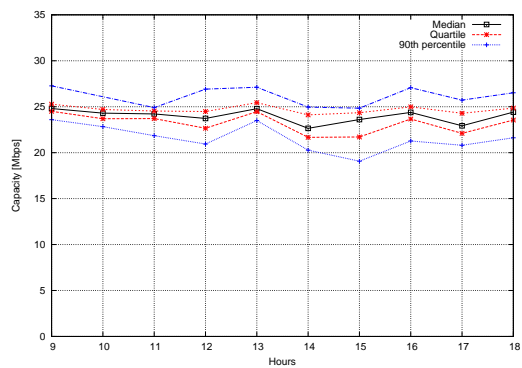}}
\quad
        \subfloat[Wi-Fi - upload]{\label{fig:wifi-back-up}\includegraphics[width=0.95\columnwidth]{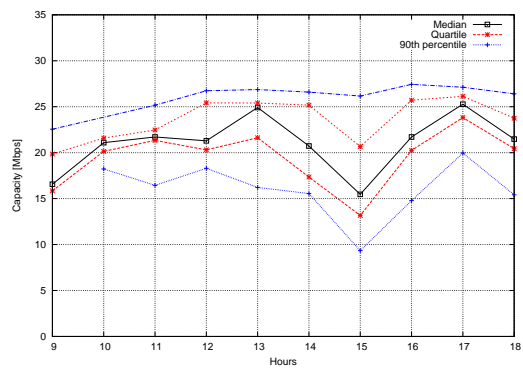}}
        \caption{Measurements in the presence of real background traffic.}\label{fig:res}
\end{figure*}

Finally, SmartProbe was tested in presence of real background traffic. In this
case, the bottleneck capacity of a Wi-Fi network was estimated several times to
assess the stability of the results. Measurements were carried out during
normal work hours, with a background traffic generated by 20-30 people.  The
experimental setup is the one shown in Figure~\ref{fig:testbed}. We performed 100
measurements every hour from 8AM to 4PM, with $n=60$, running the tool with
variable network loads. Results are depicted in Figure~\ref{fig:res} in terms
of median, inter-quartile and 90th percentile. Results are quite stable around
the median value, showing that SmartProbe is not strongly affected by the
different levels of traffic encountered during the period considered.

\section{Mapping the performance of mobile broadband operators with crowdsourcing}
\label{sec:mapping}

\begin{figure}[t]
        \centering
        \includegraphics[width=0.9\columnwidth]{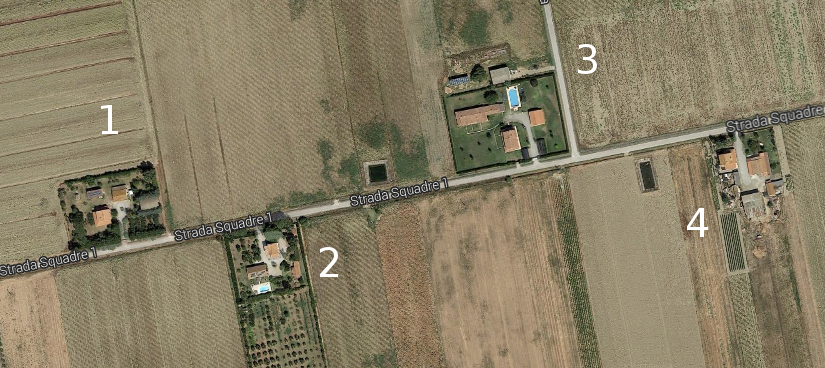}
        \caption{Mapping the performance of mobile broadband operators with crowdsourcing: considered area.}
        \label{fig:area}
\end{figure}

To demonstrate the use of SmartProbe in a crowdsourcing scenario, we
implemented an application that maps the performance of mobile broadband operators
with the help of the masses. Every time a user estimates the bottleneck
capacity, the results are georeferenced and transferred to the main server,
where they are saved onto persistent storage. When a large number of
measurements are collected, these can be used to build a map that shows the
real performance of mobile operators in relation to user positions. This
information is obviously useful for future subscribers, as they may be
interested in the real performance of operators in the area where they spend
most of their time.

\begin{figure}[t]
        \includegraphics[width=1.05\columnwidth]{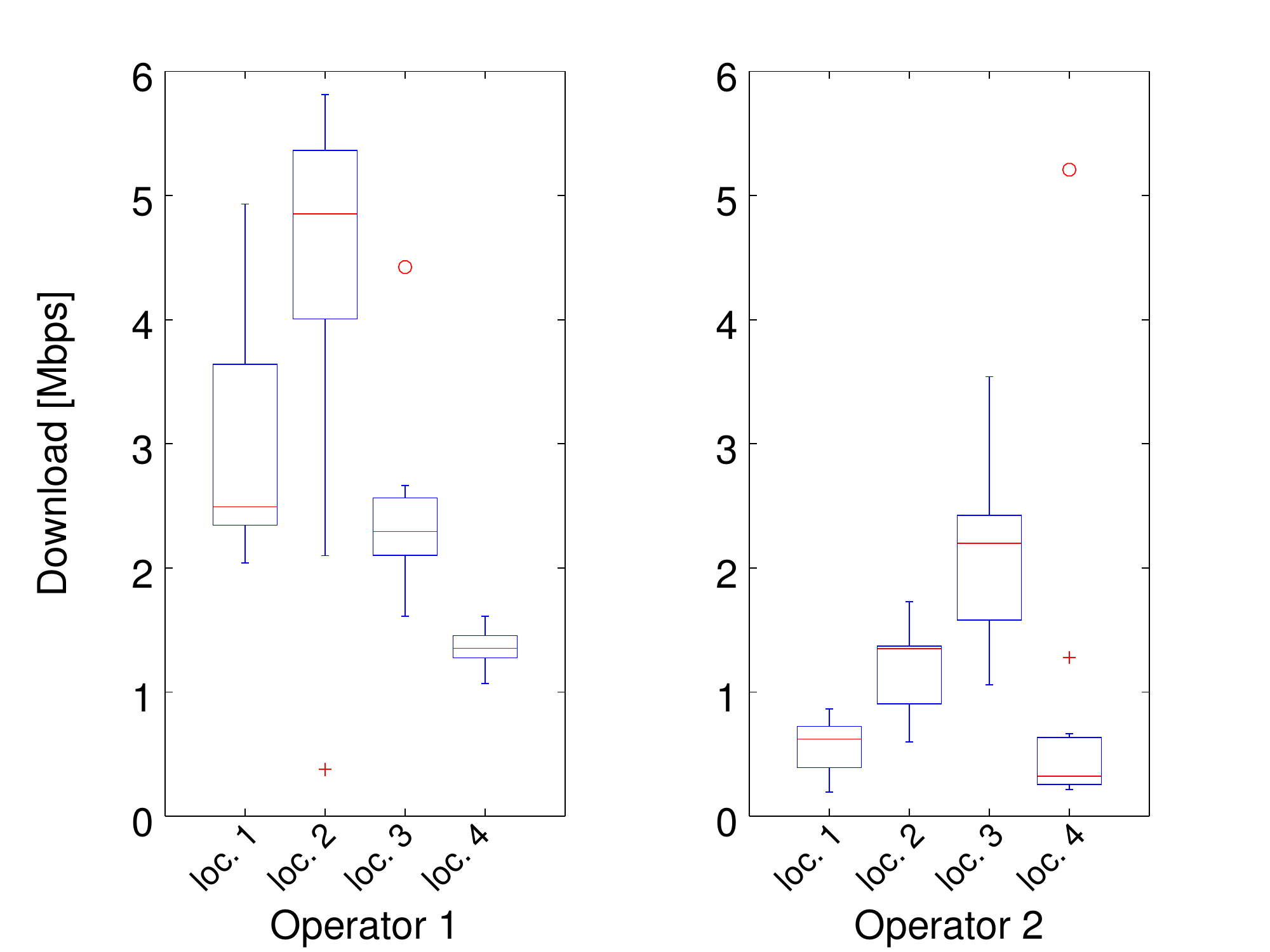}
        \caption{Mapping the performance of mobile broadband operators with crowdsourcing: results.}
        \label{fig:all_in_one}
\end{figure}

We used this proof-of-concept application to estimate the bottleneck capacity
at different locations in a suburban area in Italy (Figure~\ref{fig:area}).
For each location the bottleneck capacity was estimated using an Android
smartphone connected to the Internet via two different mobile broadband
operators (two of the major operators available in Italy, here indicated simply
as ``Operator 1'' and ``Operator 2''). In addition, for each operator the
measurement was repeated ten times. 
Figure~\ref{fig:all_in_one} shows the performance of the two operators (only
downlink for the sake of brevity) at the four selected locations (which
correspond to four houses).  In all cases, Operator 1 provides the best
performance. This demonstrates that if a measurement is collected at a given
location, the result is reasonably stable and  can be useful for all possible
future subscribers who live in the surrounding areas.

\section{Long term evaluation}

\begin{figure}[t]
  \centering
  \includegraphics[width=1.0\columnwidth]{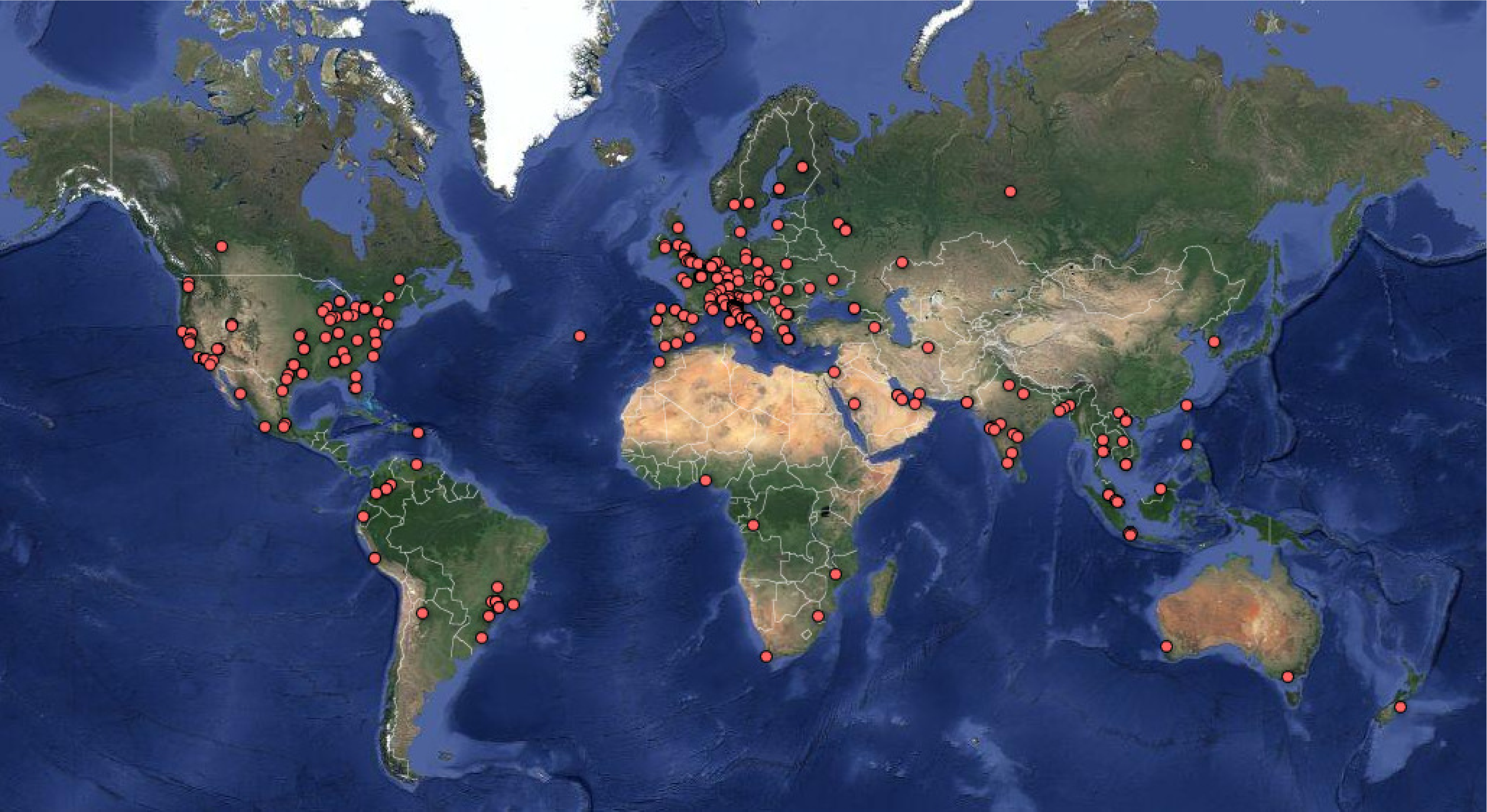}
  \caption{Location of SmartProbe users.}
  \label{fig:world_map}
\end{figure}

At the time of writing, SmartProbe has been operational for approximately one
year. The geographical distribution of SmartProbe users is depicted in
Figure~\ref{fig:world_map} (more in detail, it represents the locations where
measurements have been carried out). The majority of users is located in Europe
and North America, with several users also in South America and Asia.  Africa
and Oceania are, on the contrary, not much represented. 
As already mentioned, results are sent to a central server where they are saved
onto persistent storage. To preserve users' privacy the ID of the device is
anonymized and no personal information is transferred to the server.

\begin{table}[t]
  \centering
  \caption{Normalized number of measurements per access technology (\%).\label{tab:norm}}{
	\begin{tabular}{lr}
		\hline
    GPRS & 0.1\\
    EDGE  & 1.1\\
    UMTS & 4.6\\
    HSUPA & 0.5\\
    HSDPA & 8.6\\
    EVDO\_A & 4.9\\
    EVDO\_B & 4.6\\
    EHRPD & 2.3\\
    WI-FI & 73.8\\
    \hline 
	\end{tabular}	
}
\end{table}

The normalized number of measurements, for the different access technologies, is
reported in Table~\ref{tab:norm}. The number of measurements carried out using
declining cellular technologies (GPRS and EDGE) is quite limited. Approximately
1/4 of the measurements have been carried out using 3G technologies and the
remaining ones concern Wi-Fi links.

\begin{figure*}[t]
        \centering
        \subfloat[Upload]{\label{fig:cdf_up}\includegraphics[width=1.1\columnwidth]{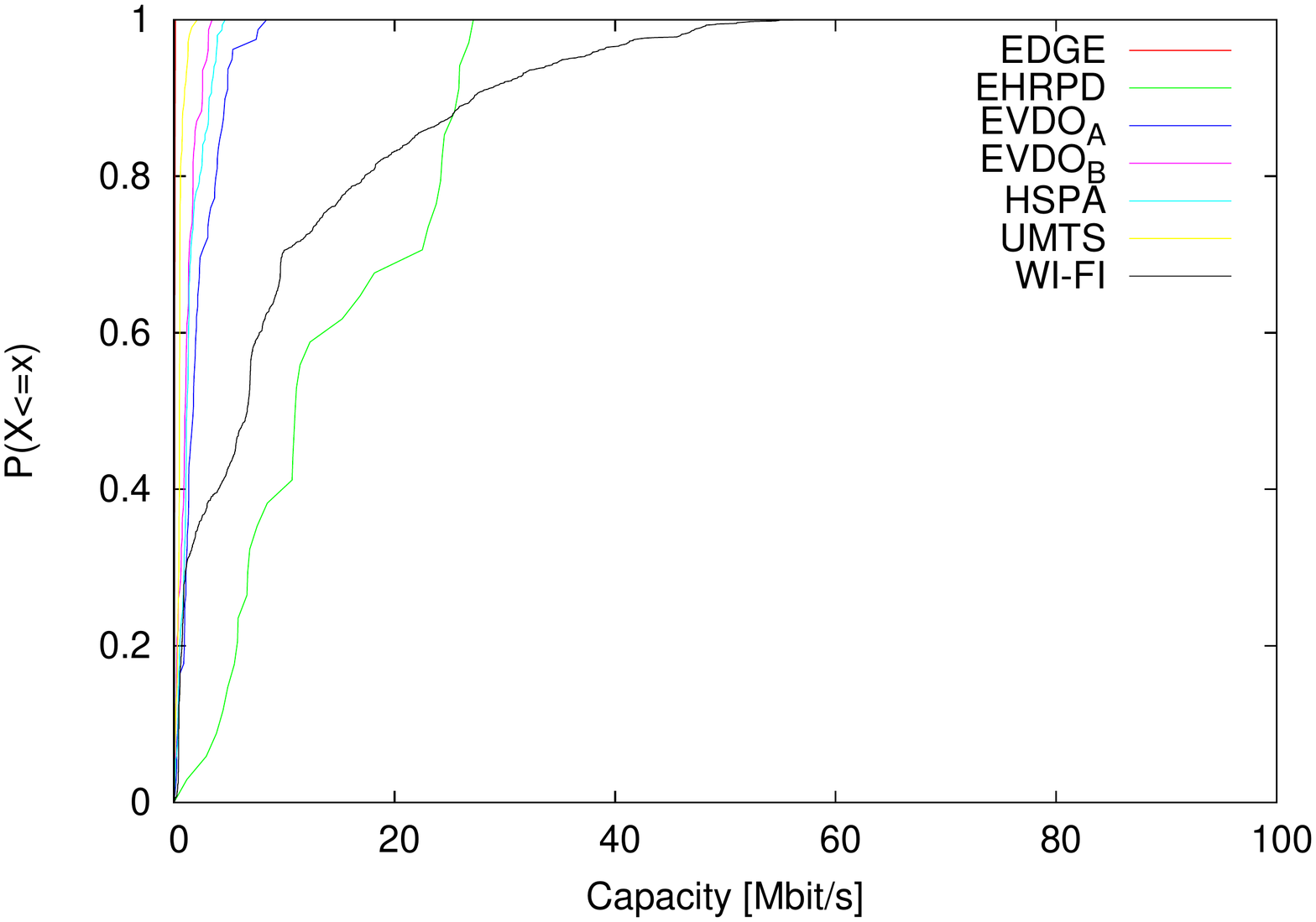}}
        \subfloat[Download]{\label{fig:cdf_down}\includegraphics[width=1.1\columnwidth]{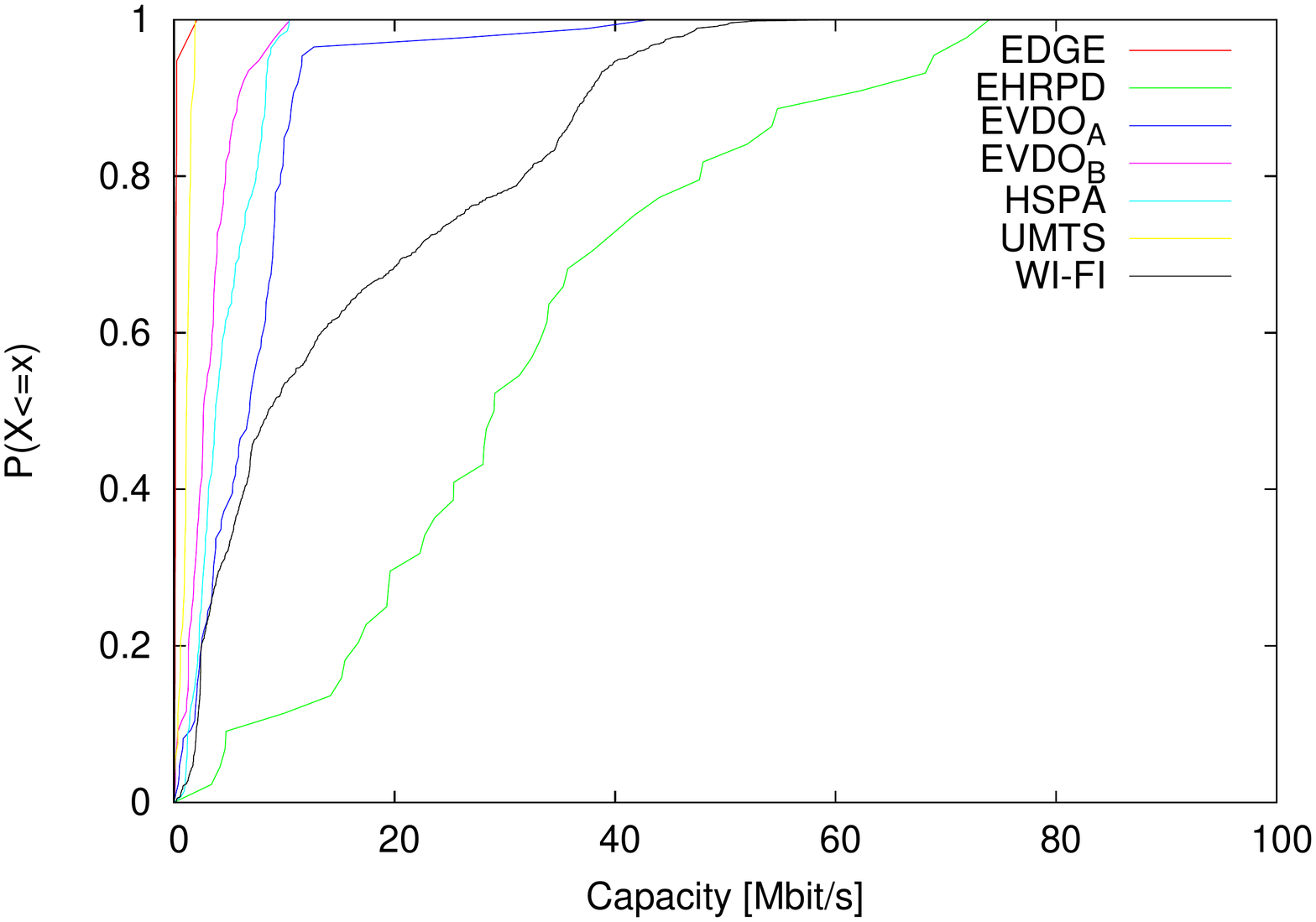}}
        \caption{Upload and download capacity (CDF) collected during approximately one year of operation.}
        \label{fig:cdf}
\end{figure*}

Figure~\ref{fig:cdf} shows the cumulative distribution function of upload and
download capacity as registered by the users of SmartProbe during the period of
operation. The total number of measurements is ~3350: ~1730 for upload and
~1620 for download. Measurements include both Wi-Fi connections and cellular
connections (these latter are grouped by access technology). As expected, 3G
technologies provide the best results for cellular connections and with EHRPD
the download capacity is even larger than Wi-Fi.  

These results must be interpreted as a first outcome of the crowdsourcing-based
technique we propose. In any case, even considering the limited period, they
clearly demonstrate that aggregating information collected by volunteers, to
provide global metrics in terms of bottleneck capacity, is feasible.
Obviously, a deeper and more detailed analysis will be possible when a larger
amount of measurements are available. It is worthwhile to remember that all
the measurements have been triggered by real users who voluntarily joined the
system. As known, increasing the number of participants in a crowdsensing
scenario is not easy, as specific incentive mechanisms have to be designed and
put into practice (but this is outside the scope of this
work)~\cite{Sun2014189}. 

\begin{figure*}[t]
        \centering
        \subfloat[Upload]{\label{fig:cdf_operators_up}\includegraphics[width=1.1\columnwidth]{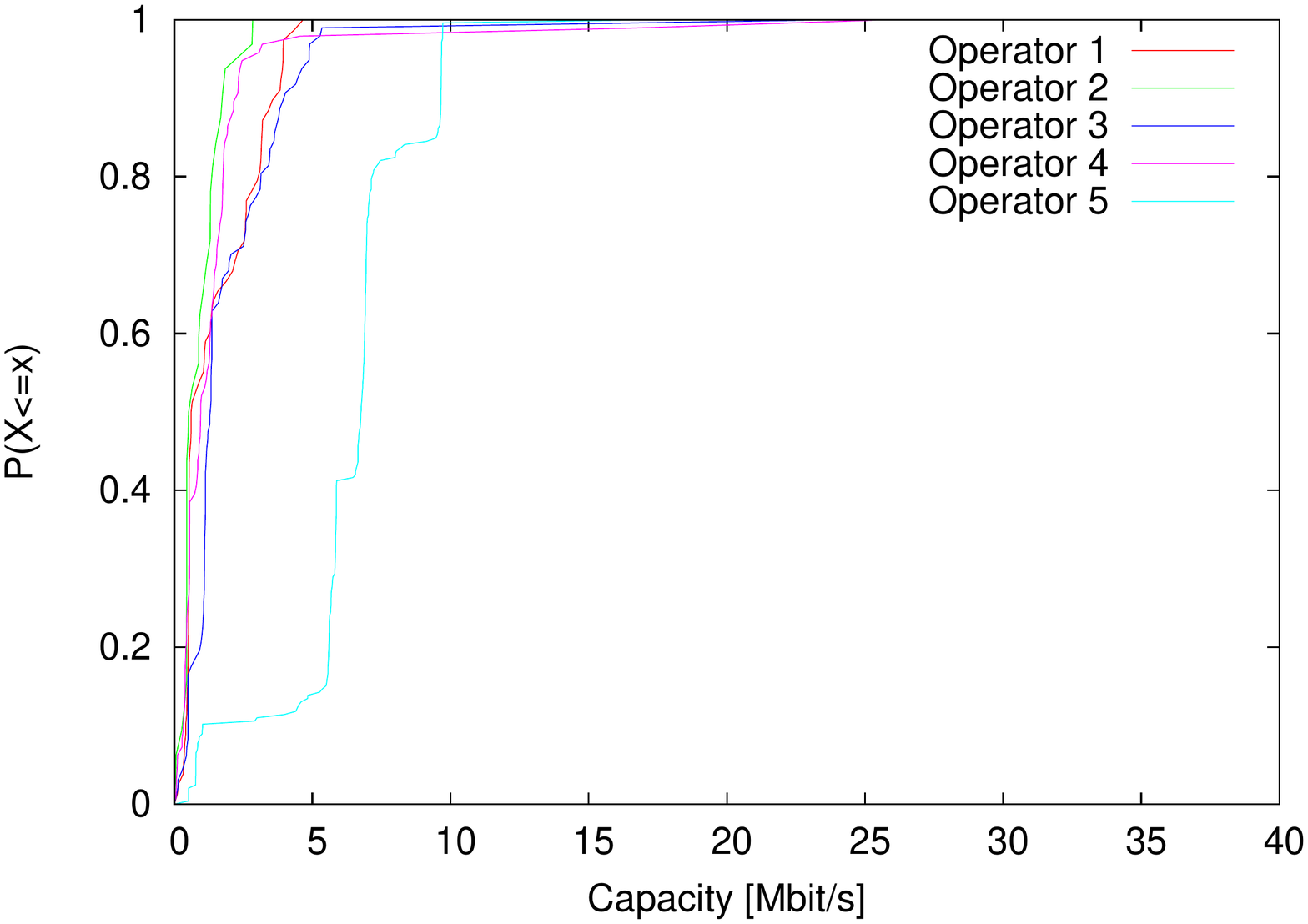}}
        \subfloat[Download]{\label{fig:cdf_operators_down}\includegraphics[width=1.1\columnwidth]{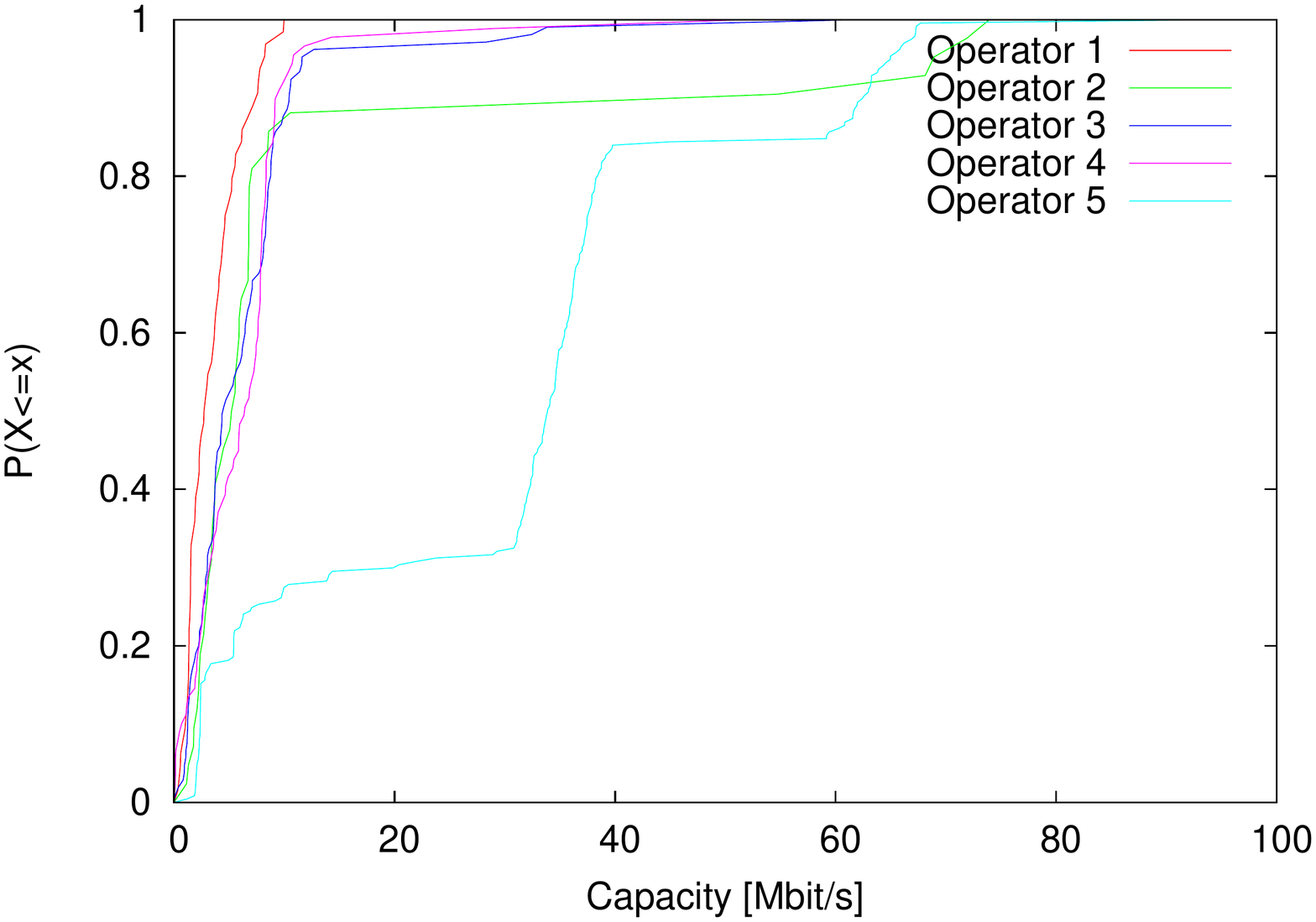}}
        \caption{Upload and download capacity (CDF) of major Italian cellular operators.}
        \label{fig:cdf_operators}
\end{figure*}

Figure~\ref{fig:cdf_operators} shows the bottleneck capacity registered when
using the five largest Italian mobile broadband operators. This kind of
analysis has been restricted to Italy because for other countries the amount of
measurements do not allow us to draw statistically sound conclusions.  As
evident, one of the operators provides significantly better performance in
terms of both upload and download capacity.  Although these results, at this
stage, cannot be used to perform a thorough comparison of cellular operators,
they already provide a first insight of the diffusion of cellular technologies
among the considered companies.  In general, a country-wide analysis like this
can be used by customers to be informed about the global performance of
operators. On the other side, an annotated map like the one presented in
Section~\ref{sec:mapping} can be extremely useful to obtain
geographic-dependent information.

\section{Related work}

This section summarizes the most significant work related to the estimation of
bottleneck capacity and to the use of crowdsourcing for network measuring.

\subsection{Estimation of bottleneck capacity}

The first tool focusing on the discovery of the bottleneck capacity was
developed by Van Jacobson in 1997. Since then, a plethora of tools have been
developed for wired networks, typically based on packet pairs (e.g.
\textit{Bprobe} \cite{Carter}, \textit{Pathrate}
\cite{Dovrolis:2004:PTC:1046014.1046015}, \textit{CapProbe} \cite{CapProbe} and
\textit{TOPP} \cite{Melander}) and packet trains (e.g. \textit{PBProbe}
\cite{Chen} and \textit{Cprobe} \cite{Carter}).  Since these tools are not
conceived for wireless environments, the results can in some cases be
inaccurate. Recently, some dedicated tools based on packet pairs have been
developed to infer the bottleneck capacity of wireless links (e.g.
\textit{WBest} \cite{Li} and \textit{AdHoc-Probe} \cite{Guang}).  However as
these tools are based on packet pairs, they can suffer from compression and
expansion phenomena. Tools based on packet trains are able to minimize the
effects of interrupt coalescence and timer resolution, but they still suffer
from contending traffic interference.  Of all these methods, PBProbe is the
best performer on wireless networks~\cite{Chen}.  However, as already stated,
the amount of traffic generated by PBProbe is not optimized, and its use on
smartphones is resource-demanding.  To the best of our knowledge, SmartProbe is
the first attempt towards a full adaptation of bottleneck capacity estimation
techniques to the smartphone platform. The only tools for smartphones
implemented so far are focused on estimating the bulk transfer capacity
\cite{Bauer}, i.e. the amount of data that can be sent between two ends via TCP
(which, as already discussed in Section~\ref{sec:theory}, is different from the
bottleneck capacity~\cite{Claffy}).

One of the first works where the problem of measuring the bandwidth is
contextualized to a wireless scenario is the one by Johnsson et
al.~\cite{johnsson2006:bandwidth}. In particular, the authors discuss the
effects caused by the probe packet size and cross-traffic on the estimation
process when 802.11 wireless links are used.

A performance assessment of four tools for capacity estimation when operating
on 802.11 links is presented in~\cite{1613635}. Experiments have been carried
out in a semi-anechoic chamber, in order to evaluate the tools in a controlled
and interference-free environment. Repeatability of procedures and the use of a
digital storage oscilloscope allowed the authors to collect extremely accurate
timestamps and to deeply study the interaction between capacity estimation
tools and network interface cards.

\subsection{Crowdsourcing for evaluating network properties}

The crowdsourcing approach has been used to evaluate and measure network
properties both in wired and wireless scenarios.

DIMES is a distributed measurement infrastructure that collects information on
the topology of the Internet and its evolution~\cite{ssh05}.  The key idea
behind DIMES is a shift from dedicated measurement architectures to a large
community of volunteers: each participant runs a monitoring agent that carries
out traceroute- and ping-based campaigns, then results are aggregated to
produce a map at the autonomous system level of abstraction.  Besides
parallelization of workload, the heterogeneity of participants in terms of
location provides the opportunity to measure the Internet from different points
of view. This is a significant improvement compared to solutions characterized by
a relatively small number of vantage points. In summary, crowdsourcing is beneficial not
only in terms of raw power, but also because of the specific nature of every
individual participant.

Dasu is an experimentation platform that supports measurement activities
at the Internet's edge on end user machines~\cite{sanchez13:dasu}. Dasu
has been implemented as an extension of a peer-to-peer client
(BitTorrent) to leverage its popularity and wide coverage. An experiment
administrator can assign tasks to the participating clients. The
measurement activities to be executed on user devices are specified
through simple ``when-then'' rules and comprise both active tools
(ping, traceroute, etc.) and passive data collectors. The availability
of a large number of cooperating users supports the execution of large
scale experiments aimed at studying routing asymmetry and Internet topology.

Similarly, crowdsourcing has been used to detect service-level network
events~\cite{choffnes}, and to characterize ISPs and evaluate their
performance~\cite{bos11}. Also in these cases, the BitTorrent client has been
adopted as the implementation platform, since P2P sessions are
network-intensive (this enables the passive collection of information) and
characterized by long sessions (thus providing extensive monitoring periods).
The unique perspective of participants and the involvement of a large number of
end-users enable the collection of an unprecedented amount of information
on the status of the network, especially at the edges of the Internet.

BSense is a system aimed at creating maps concerning the quality of broadband
connections~\cite{Bernardi:2014:BFO:2710113.2710126}. The system relies on a
software agent that is executed on broadband users' machines. The agent
periodically measures the latency, packet loss rate, and bandwidth of the
broadband connections, then it uploads the obtained results onto the BSense
server. The framework has been tested with the help of 60 volunteers located
in a Scotland region. Differently from SmartProbe, BSense is not specifically
designed for estimating capacity in a wireless scenario: the default
configuration for the download and upload sessions uses 8 concurrent UDP/TCP
flows at 400 packets/s, with 1024 byte packets.  

A large study about broadband performance is presented
in~\cite{Sundaresan:2011:BIP:2018436.2018452}: latency, packet loss, and
throughput measurements have been collected from nearly 4000 users and across
16 ISPs in USA. The largest part of monitored homes relies on gateways
specifically deployed for studying network performance from the point of view of
residential users.  In particular, running experiments from such vantage points
enables fine grained control on confounding factors, such as cross traffic or
home wireless networks.

Hobbit is a measurement platform for broadband monitoring
\cite{hobbit2014:donato}. The platform includes measurements clients executed
on users' machines, measurement servers that cooperate with clients, and a
management server for coordinating activities, planning experiments, and
collecting results. A deployment of 400 clients has been used to study the
performance of a set of ISPs in terms of fixed broadband access.

We believe that SmartProbe represents a significant addition to the landscape
depicted above: it shares with these systems the key idea of using people as a
means for evaluating the properties of large-scale networks with limited
efforts, and expands existing work with bottleneck estimation in a wireless
scenario.

Cognitive radios (CRs) are intelligent wireless communication systems that
adapt their behavior to changing spectral conditions~\cite{1391031}. By
automatically adapting their parameters of operation to the network status, CRs are
able to more efficiently use the radio spectrum and to provide reliable
communication~\cite{5198712}.
In~\cite{zhao07:maps}, the authors discuss the use of radio environmental maps
(REMs) for cognitive wireless regional area networks. REMs operate as an
integrated database that characterizes the radio environment using geographical
information, spectral regulations, location and activities of radios, and
policies.  REMs are updated using observations produced by CR nodes, which in
turn receive relevant information through the CR network itself.  To a certain
extent, part of the content of REMs is generated according to the crowdsourcing
paradigm.  Similar ideas are also discussed in~\cite{Zhao2006337,4286322}. 

At first glance, these maps and the one produced by SmartProbe could be
considered as similar. However, a more extensive analysis makes it clear that
the purpose of REMs and SmartProbe maps is totally different and that the two
systems operate on completely separate levels. In fact, REMs provide
information used to tune the operational parameters at the lower levels of the
networking stack (mostly the physical layer and the data link layer); the aim
is to improve performance or to make efficient use of the spectrum.  In
contrast, the system we propose produces information dedicated to the user
level and is completely independent of the underlying communication
technologies and protocols.  Unlike CRs, SmartProbe does not improve
communication efficiency; instead it provides answers to questions like ``which
cellular operator performs best in an area of interest?''. Answers are provided
using information generated by other users via crowdsourcing.

\section{Conclusions}

The mobile crowdsensing paradigm can be extremely useful for analyzing the
characteristics of large-scale networks. With the help of the masses it is now
possible to collect an unprecedented amount of information on the performance
of networked systems. At the same time, the mobility of users and devices makes
it possible to analyze large networks from a geographical point of view.  However,
these opportunities come at the cost of increased technical complexity.
Smartphones are resource-constrained devices, especially from the point of view
of energy and communication costs, and thus specific energy- and
bandwidth-saving techniques need to be designed and put into practice.
Protocols need to cope with a possibly large number of users and the back-end
infrastructure has to store and process huge amounts of data.

We have described the design and implementation of a mobile crowdsourcing
system aimed at measuring the bottleneck capacity of Internet paths. SmartProbe
generates less traffic than similar tools, in order not to compromise the user
experience, and includes server-side mechanisms to support simultaneous
measurement requests.  The presented application demonstrates that a collective
and georeferenced evaluation of the performance of mobile broadband operators
can be used to support intelligent user decisions.

\section*{Acknowledgments}
This work has been partially supported by the European Commission within the
framework of the CONGAS project FP7-ICT-2011-8-317672 and by the University of
Pisa within the ``Metodologie e Tecnologie per lo Sviluppo di Servizi
Informatici Innovativi per le Smart Cities - PRA 2015'' project.

\newpage
\onecolumn
\footnotesize
\bibliographystyle{elsarticle-num} 
\bibliography{all-references}

\end{document}